\documentclass[twocolumn, 10pt]{article}
\usepackage[utf8]{inputenc}
\usepackage[T1]{fontenc}
\usepackage{mathptmx}
\usepackage{amsmath, amssymb}
\usepackage{graphicx}
\usepackage{subcaption}
\usepackage{float}
\usepackage{bm}
\usepackage{dcolumn}
\usepackage{stfloats}  
\usepackage[margin=2cm, columnsep=0.6cm]{geometry}
\usepackage[numbers, sort&compress]{natbib}
\newenvironment{acknowledgments}
  {\section*{Acknowledgments}}
  {}
\usepackage{booktabs}
\usepackage{hyperref}
\usepackage{xurl}  
\usepackage{tabularx}
\usepackage{authblk}

\usepackage{bm}
\DeclareMathVersion{light}
\SetSymbolFont{operators}{light}{OT1}{cmr}{m}{n}
\SetSymbolFont{letters}{light}{OML}{cmm}{m}{it}
\SetSymbolFont{symbols}{light}{OMS}{cmsy}{m}{n}
\newcommand{\mbm}[1]{\mathversion{light}\bm{#1}\mathversion{normal}}
\usepackage{authblk}

\usepackage{abstract}

\usepackage[raggedright]{titlesec}  

\begin{document}

\title{Three-dimensional density and air-rock interface reconstruction with muography: Application to the TianQin tunnel}

\author[1]{Songran Qi}
\author[1]{Tao Yu}
\author[1]{Shihan Zhao}
\author[1]{Yunsong Ning}
\author[1]{Aiyu Bai}
\author[1]{Yu Chen}
\author[1]{Yi Yuan}
\author[1]{Mingchen Sun}
\author[1]{Zhirui Liu}
\author[1]{Liang Xian}
\author[1]{Hengye Xu}
\author[1]{Hao Jiang}
\author[1]{Zhichao Wang}
\author[2]{Shuhang Zhang}
\author[3]{Su Zhan}
\author[1,$\dagger$]{Jian Tang}
\affil[1]{School of Physics, Sun Yat-sen University, 510275 Guangzhou, China}
\affil[1]{Platform for Muon Science and Technology, Sun Yat-sen University, Guangzhou, China}
\affil[2]{School of Geospatial Engineering and Science, Sun Yat-sen University}
\affil[3]{TianQin Research Center for Gravitational Physics, Sun Yat-sen University}
\affil[$\dagger$]{Corresponding authors: tangjian5@mail.sysu.edu.cn}
\date{\today}

\twocolumn[
  \begin{@twocolumnfalse}
    \maketitle
    \begin{abstract}
      \noindent
      Muography is a non-invasive imaging technique that uses cosmic-ray muons, commonly divided into transmission (absorption) and scattering muography. For transmission muography, the inversion algorithm critically determines reconstruction quality. However, widely used schemes may produce smearing artifacts when measurement locations are limited and data are sparse. We develop an optimized Metropolis–Hastings (M–H) algorithm that mitigates smearing and retrieves sharper, more accurate density distributions without auxiliary data. Additionally, we implement an inverse distance weighting (IDW) approach to reconstruct the air–rock interface from muon measurements. The optimized M–H algorithm is applied in Monte Carlo simulations and applied to field data from the TianQin Tunnel experiment using the MuGrid-v2 detector. The IDW-reconstructed air–rock interface is validated against Light Detection and Ranging (LiDAR) measurements. In simulations, the optimized M–H algorithm improves high-density anomaly detection precision from $42\%$ to $100\%$ at threshold $5.1\,\mathrm{g/cm^3}$, with gains of $6\%$ to $42\%$ across other threshold and low-density scenarios, together with the TianQin Tunnel reconstructions, these results demonstrate the effectiveness of the proposed approach.
    \end{abstract}
    \vspace{0.5cm}
  \end{@twocolumnfalse}
]


\section{Introduction}
\label{sec:intro}
Muography originated from the 1950s with E.P. George's pioneering work on cosmic-ray muons to measure rock overburden\cite{pCosmicRaysMeasure1955}, and was further developed for large-scale imaging in the 1990s\cite{nagamineMethodProbingInnerstructure1995}. The technique exploits cosmic-ray muons as natural probes and operates in two modes: transmission muography measures flux attenuation governed by the Bethe–Bloch equation\cite{bethe1930theorie} to reconstruct density distributions and air-rock interfaces, while scattering muography analyzes Coulomb scattering angles—related to atomic number $Z$\cite{borozdin2003RadiographicImaging}—to achieve finer spatial resolution but is typically limited to small or medium sized targets due to angular measurement requirements. Alessandro Lechmann presents a comprehensive and detailed guide for it\cite{lechmann2021MuonTomography}. 

This paper focuses on transmission muography as it is more suitable for large-scale target imaging and many previous studies have demonstrated the versatility of it across volcanology, geology, and archaeology. Carbajal et al. reconstructed the three-dimensional density distribution of La Soufrière's lava dome using least-squares optimization\cite{rosas-carbajal2017ThreedimensionalDensity}. L'Istituto Nazionale di Fisica Nucleare (INFN) identified hidden cavities in geological surveys via clustering algorithms\cite{cimmino3DMuographySearch2019}, and Lanzhou University applied iterative global optimization to image the internal structure of Xi'an's ancient defensive wall\cite{liuHighprecisionMuographyArchaeogeophysics2023}. Additionally, muography can reconstruct air–rock or ice–rock interfaces, which is particularly valuable where traditional geophysical methods are hindered by thick ice cover, dense vegetation, steep terrain, or inaccessible locations. In such scenarios, muography provides an independent dataset for interface detection without requiring direct surface access. For instance, muography has been used to measure ice-bedrock interfaces in Central Swiss Alps\cite{nishiyamaFirstMeasurementIcebedrock2017, nishiyamaBedrockSculptingActive2019}. 

The reconstruction process is critical in transmission muography, as it involves solving an inverse problem whose solution directly determines the reconstruction quality. Several algorithms have been developed for this purpose, including the damped least squares method\cite{wampler1986ManipulatorInverse, nakamura1986InverseKinematic}, Limited-memory Broyden–Fletcher–Goldfarb–Shanno (L-BFGS)\cite{byrdLimitedMemoryAlgorithm1995}, and Simultaneous Algebraic Reconstruction Technique (SART)\cite{kak2002PrinciplesComputerized}. These classical methods have been successfully applied in muography studies\cite{tanakaThreedimensionalComputationalAxial2010, procureur3DImagingNuclear2023, liuDeepInvestigationMuography2024}. However, they may produce unsatisfactory results with severe smearing issue when measurement data are sparse or detector locations are limited. More recently, several specialized methods have been developed for applications, including integrated processing of muography and gravity anomaly data\cite{nishiyamaIntegratedProcessingMuon2014, nishiyama3DDensityModeling2017}, process with the seed algorithm\cite{liuDeepInvestigationMuography2024}, and process with the back-projection technique\cite{borselli2022ThreedimensionalMuon}.
While these methods demonstrate promising results, they require additional constraints: the seed algorithm necessitates manual specification of initial seed points. The integrated approach relies on auxiliary gravity data as prior information, and the back-projection technique can not reconstruct accurate density distribution. These requirements limit their applicability in scenarios where such prior knowledge is unavailable or density distribution is needed.

In this paper, we develop an optimized Metropolis–Hastings (M-H) algorithm that reconstructs density distributions without manual initialization or auxiliary data. This algorithm also completes our team's imaging pipeline by providing a robust reconstruction framework to complement the existing in-house-developed muon scintillator detector, MuGrid-v2\cite{yu2025MuGridv2Novela}. The algorithm addresses the smearing problem in sparse measurement scenarios. Additionally, we apply the Inverse Distance Weighting (IDW) method to reconstruct air-rock interface. To test the capability of optimized M-H algorithm, we test it in a Monte Carlo simulation. We then deploy the MuGrid-v2 detector inside the Tianqin Tunnel to image its internal structure and the overlying rock-air interface. To evaluate the IDW-based elevation reconstruction, we compare our results with interface data acquired by Light Detection and Ranging (LiDAR). The section below introduces the working principle of muography and its reconstruction method.


\begin{figure*}[htbp]
  \centering
    \includegraphics[width=0.9\linewidth]{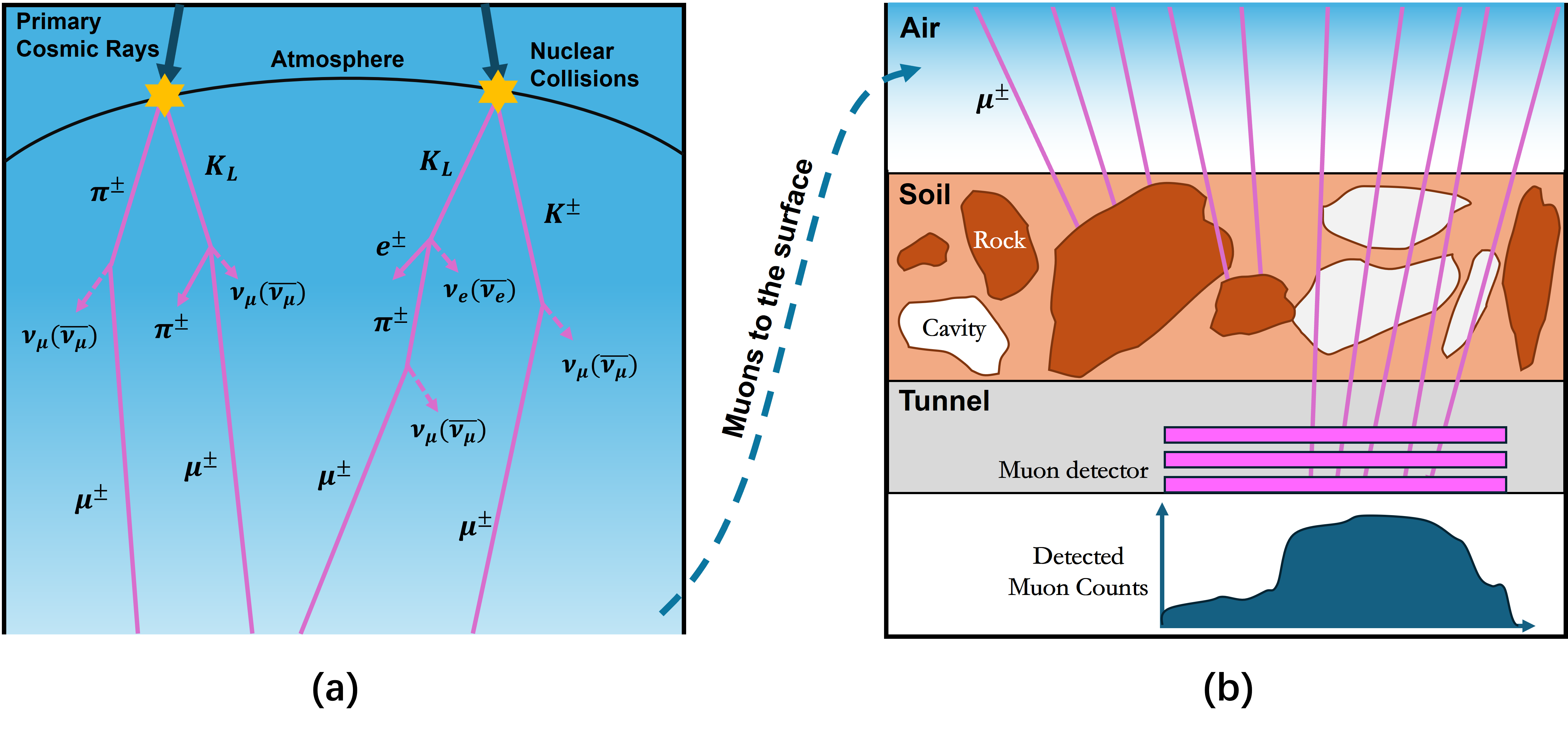}
  \caption{Principles of cosmic-ray muon generation and transmission muography. (a) Cosmic rays collide with atmospheric nuclei, producing pions and kaons that decay into muons. (b) Muons penetrate the overburden and are detected in a tunnel through muon detector. The two-dimensional distribution below the detector illustrates the transmission muography principle: fewer muons are detected along high-density paths, while more muons traverse low-density regions.}
  \label{c2:fig1}
\end{figure*}

\section{Overview of cosmic-ray muon imaging}
\label{sec:ImageMethod}
Cosmic-ray muons serve as the probe particles in muography. They are produced through nuclear collisions between cosmic rays (primarily protons and nuclei) and atmospheric nuclei, generating pions and kaons that subsequently decay into muons (Fig.~\ref{c2:fig1}a). With typical energies around $4\,\mathrm{GeV}$ at the peak of the momentum distribution\cite{particledatagroup2018ReviewParticle}, cosmic-ray muons can penetrate hundreds of meters of rock. This penetrating capability enables transmission muography to image the internal structure of target objects by measuring muon flux attenuation. As illustrated in Fig.~\ref{c2:fig1}b, muons traversing high-density rock are more likely to be absorbed, resulting in lower flux, whereas those passing through low-density cavities maintain higher flux, this differential attenuation forms the basis of transmission imaging. The energy loss process of muons in matter is described by the Bethe formula\cite{bethe1930theorie,particledatagroup2018ReviewParticle}:

\begin{equation}
    \frac{\mathrm{d} E}{\mathrm{d} x}   = Kz ^{2}\frac{Z}{A} \frac{1}{\beta ^{2} } \left [ \frac{1}{2} \ln_{}({\frac{2m_{e}c^{2} \beta ^{2}\gamma^{2}W_{max}   }{I^{2} })-\beta ^{2}-\frac{\delta (\beta\gamma)}{2} }  \right ],
    \label{c3:eq10}
\end{equation}

where $K$ is a constant, $z$ is the charge number of the incident particle ($z=1$ for muons), $m_e$ and $c$ represent the electron mass and speed of light, respectively. $Z$ and $A$ denote the atomic number and atomic mass number of the medium, respectively, with $I$ being the mean excitation energy of the medium atoms. $W_{\mathrm{max}}$ is the maximum kinetic energy transferred in a single muon-atom collision, and $\delta$ accounts for the density effect correction. $\beta$ is defined as the ratio of particle velocity to the speed of light $c$, and $\gamma = 1/\sqrt{1-\beta^2}$. $X$ represents the opacity, which is defined in Eq.~(\ref{c3:eq5}): 

\begin{eqnarray}
X = \int_{\text{path}} \rho(L)\,dL,
\label{c3:eq5}
\end{eqnarray}

where $\rho(L)$ is the density along the muon trajectory, and $L$ is the path length that the muon has traversed. This integral relationship is bidirectional: it enables density reconstruction from measured opacity (inverse problem), and also allows path length estimation for air-rock interface reconstruction given density distribution (forward problem). 

For density reconstruction, by inverting and integrating Eq.~(\ref{c3:eq10}), the opacity along each line of sight from the detector can be obtained. This opacity can either be directly converted into a two-dimensional average density distribution or serve as input for further three-dimensional reconstruction. To calculate the opacity, the muon survival rate $S$ is obtained by comparing the attenuated muon flux through the target with the reference open-sky flux:

\begin{eqnarray}
S = \frac{N_{u} (\theta)\Delta T_{os} }{N_{os} (\theta)\Delta T_{u}}=\frac{ \int_{E_{min}(X) }^{\infty}\Phi(\theta, E)dE  }{\int_{E_{0}(X) }^{\infty}\Phi(\theta, E)dE},
\label{c3:eq6}
\end{eqnarray}

where $N_{u}$ and $N_{os}$ represent the number of muons detected beneath the target object and under the open sky respectively. $\Delta T_{u}$ and $\Delta T_{os}$ are the effective data collection times in the corresponding cases. $\Phi(\theta, E)$ is the muon flux at zenith angle $\theta$ and muon energy $E$. $E_{0}$ and $E_{min}$ are the minimum muon energies required to trigger the detector under the open sky and beneath the target object, respectively.

By measuring the muon survival rate $S$ from open sky and underground data, combined with a differential muon flux model $\Phi(\theta, E)$, the minimum energy $E_{\min}$ can be inferred, see Appendixes~\ref{sec:Appen} for details. The differential flux model is adopted from EcoMug\cite{paganoEcoMugEfficientCOsmic2021}.

In our experiment, the relationship between $E_{min}$ and $X$ for different materials is obtained with linear interpolation from a table of $E_{min}$ and the corresponding Continuously Slowing Down Approximation (CSDA) range. These tables are obtained from the Particle Data Group (PDG) database\cite{olive2014ReviewParticle, PDGLive2015}.

After acquiring opacity $\mbm{X}$, both three-dimensional density reconstruction and interface reconstruction is possible with Eq.~(\ref{c3:eq1}):

\begin{eqnarray}
\mathbf{L}\mbm{\rho} = \mbm{X},
\label{c3:eq1}
\end{eqnarray}

where $\mbm{L}$ is a matrix representing the path lengths in different directions, $\mbm{X}$ is the opacity vector, and $\mbm{\rho}$ is the density vector. Given the measured opacity $\mbm{X}$ and assuming homogeneous density $\mbm{\rho}$, we can solve for $\mbm{L}$ to reconstruct an air-rock interface. Conversely, if the path lengths $\mbm{L}$ are known from previous air-rock interface measurement and $\mbm{X}$ is measured, the three-dimensional density distribution $\mbm{\rho}$ can be obtained by solving the inverse problem. The following sections describe these two reconstruction methods in detail.

\section{Reconstruction method}
\label{sec:ReMethod}
\subsection{Air-rock interface reconstruction method}
\label{sec:TReMethod}
\begin{figure*}[htbp]
  \centering
  \begin{subfigure}[b]{0.75\textwidth}
    \centering
    \includegraphics[width=\textwidth]{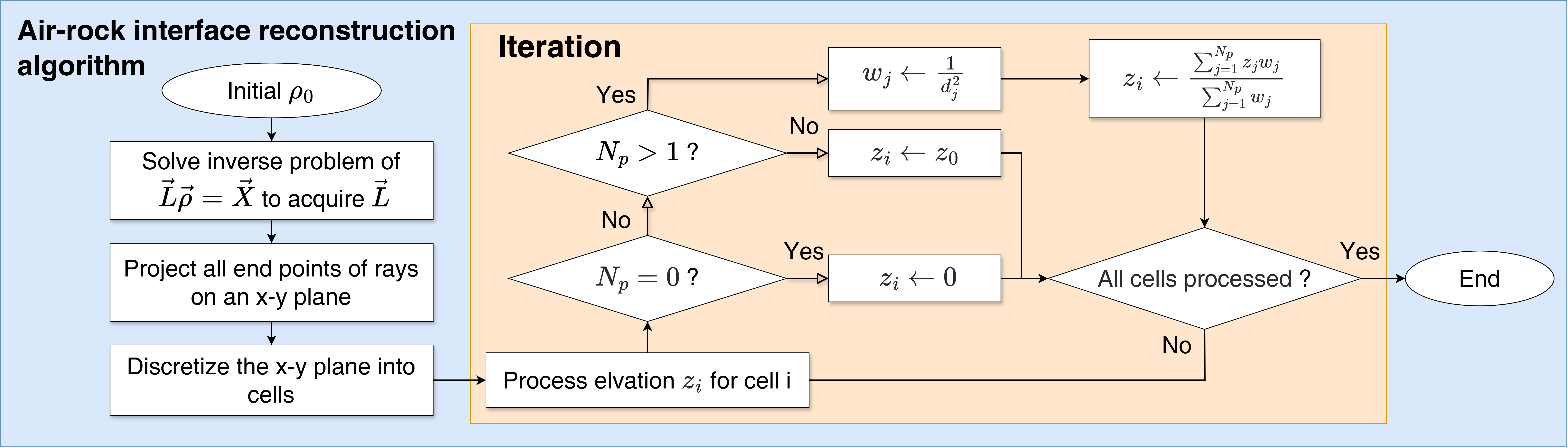}
    \caption{Air-rock interface reconstruction algorithm.}
    \label{c3:fig:terrain}
  \end{subfigure}
  \hfill
  \begin{subfigure}[b]{0.75\textwidth}
    \centering
    \includegraphics[width=\textwidth]{pic/chap03/MH_MIX_widerVer.drawio.pdf}
    \caption{Optimized M-H algorithm}
    \label{c3:fig:mh}
  \end{subfigure}
  \caption{Work flows of the reconstruction algorithms.}
  \label{c3:fig:algorithms}
\end{figure*}
As described above, a two-dimensional elevation map (air-rock interface) can be calculated through Eq.~(\ref{c3:eq1}). The detailed processing steps are as follows:

\begin{enumerate}
    \item Provide the average rock density (if assuming uniform rock density distribution), or a prior density distribution $\rho_{0}$ to solve the inverse problem of $\mathbf{L}\mbm{\rho} = \mbm{X}$ for acquiring $\mbm{L}$. Each element $L_{i}$ in $\mbm{L}$ represents the length of a ray emitted from the detector in different directions.
    
    \item Project endpoints of the rays in $\mbm{L}$ on an x-y plane with their corresponding elevation. Then discretize the x-y plane into cells. The cells resolution determines the accuracy of the elevation map.

    \item Traverse all grid cells in the discretized plane. If the number of points within a cell $N_{p}$ satisfies $N_{p}=0$, its elevation is assigned as $0$. Otherwise if $N_{p}$ satisfies $N_{p}=1$ the end point elevation is assigned as the cell elevation. Alternatively if the $N_{p}$ satisfies $N_{p}>1$, the IDW method is applied to calculate the weighted average elevation of this cell. Eq.~(\ref{c3:eq2}) shows how the IDW method works:

    \begin{eqnarray}
    z=\frac{\sum_{i=1}^{N_{p}} z_{i}w_{i}}{\sum_{i=1}^{N_{p}} w_{i}}\;,
    \label{c3:eq2}
    \\
    w_{i}=\frac{1}{d_{i}^2}.
    \label{c3:eq3}
    \end{eqnarray}

    Here, $z$ is the weighted average elevation, $z_{i}$ represents the elevation of the $i$-th end point within the cell, and $w_{i}$ is the weights of each point, defined as the reciprocal of the square of its distance to the cell center $d_{i}$. 

\end{enumerate}

Fig.~\ref{c3:fig:algorithms} (a) summarizes the processing steps and illustrates the flowchart of the air-rock interface reconstruction method. 


\subsection{Density distribution reconstruction method}
\label{sec:DReMethod}
For three-dimensional density distribution reconstruction, we first obtain the path length matrix $\mbm{L}$ through predefined voxels between the detector and the topographic surface using LiDAR-derived air-rock interface models. With $\mbm{L}$ determined, we can solve the inverse problem in Eq.~(\ref{c3:eq1}) using the optimized Metropolis-Hastings (M-H) method, which is essentially a Markov Chain Monte Carlo (MCMC) approach that generates samples iteratively and uses their ensemble average as the final solution. Since this method requires an initial guess, we employ the results from L-BFGS-B (bound-constrained L-BFGS) and SART methods as the initial sample. Specifically, we use the SciPy implementation of L-BFGS-B\cite{zhuAlgorithm778LBFGSB1997, moralesRemarkAlgorithm7782011, virtanen2020SciPy10} and our own implementation of SART.

The detailed steps of the optimized M-H reconstruction method are as follows. 
\begin{enumerate}
    \item Initialize the current sample. In our case, we use the sample calculated from the L-BFGS-B algorithm and the SART algorithm.
    
    \item Apply a $3\times 3$ Gaussian convolution kernel to the current sample. If part of the kernel extends beyond the sample boundary, re-normalize the kernel based on the remaining portion before application. This step generates a vector $\mbm{G}$ where each element in $\mbm{G}$ represents the density of the corresponding cubic voxel after Gaussian filtering.
    
    \item Use the vector $\mbm{G}$ and a predefined fixed bias rate $b$ to set the iteration step size for each element. Hence, each current sample element $\rho^{i}_{j}$ is randomly perturbed within the interval $[-b G^{i}_{j}, b G^{i}_{j}]$, where $i$ is the iteration count and $j$ denotes the voxel index in the sample vector. The perturbed vector is marked as the new sample $\mbm{\rho'}$. To ensure that all values in the new sample have physical meaning, all negative values will be set to zero in $\mbm{\rho'}$.
    
    \item Calculate the likelihoods of the current sample $\mbm{\rho}$ and the new sample $\mbm{\rho'}$, denoted as $L_{c}$ and $L_{d}$ respectively, using Eq.~(\ref{c3:eq4}):

    \begin{eqnarray}
    L_{d}=\exp\left[-\frac{1}{2T_{s}} \sum_{i=1}^{I}(|X_{i}|-|X_{i}^{\text{the}}|)^{2}\right].
    \label{c3:eq4}
    \end{eqnarray}

    Here, $X_{i}$ is the opacity measured experimentally, $X_{i}^{the}$ is the opacity calculated from the input sample using Eq.~(\ref{c3:eq1}), and $T_{s}$ is the system temperature.
    
    \item Compare $L_{c}$ and $L_{d}$. If $L_{d}$ is larger than $L_{c}$, update the current sample to the new sample. If $L_{d}$ is smaller than $L_{c}$, the current sample is updated with probability $\frac{L_{d}}{L_{c}}B_{p}$, where $B_{p}$ is a acceptance rate satisfying $B_{p} < 1$. If the update is rejected, return to step 2.
    
    \item Increment the accept counter $I$. If the number of accepted sample reaches the interval $I_{s}$, add the current sample to the posterior sample set $F$ and reset the counter.
    
    \item Check whether the size of the posterior samples, $N_{f}$, has reached the target size $N_{s}=N_{b}+N_{n}$. where $N_{b}$ is the number of burn-in samples and $N_{n}$ is the number of effective samples required. If $N_{f} = N_{s}$, proceed. Otherwise, return to step 2.

    \item Discard the first $N_{b}$ burn-in samples and compute the final density distribution as the element-wise mean of the remaining 
$N_{n}$ samples.
\end{enumerate}

Fig.~\ref{c3:fig:algorithms} (b) shows how the optimized M-H algorithm works.

\begin{figure*}[htbp]
    \centering
    \includegraphics[width=0.6\textwidth]{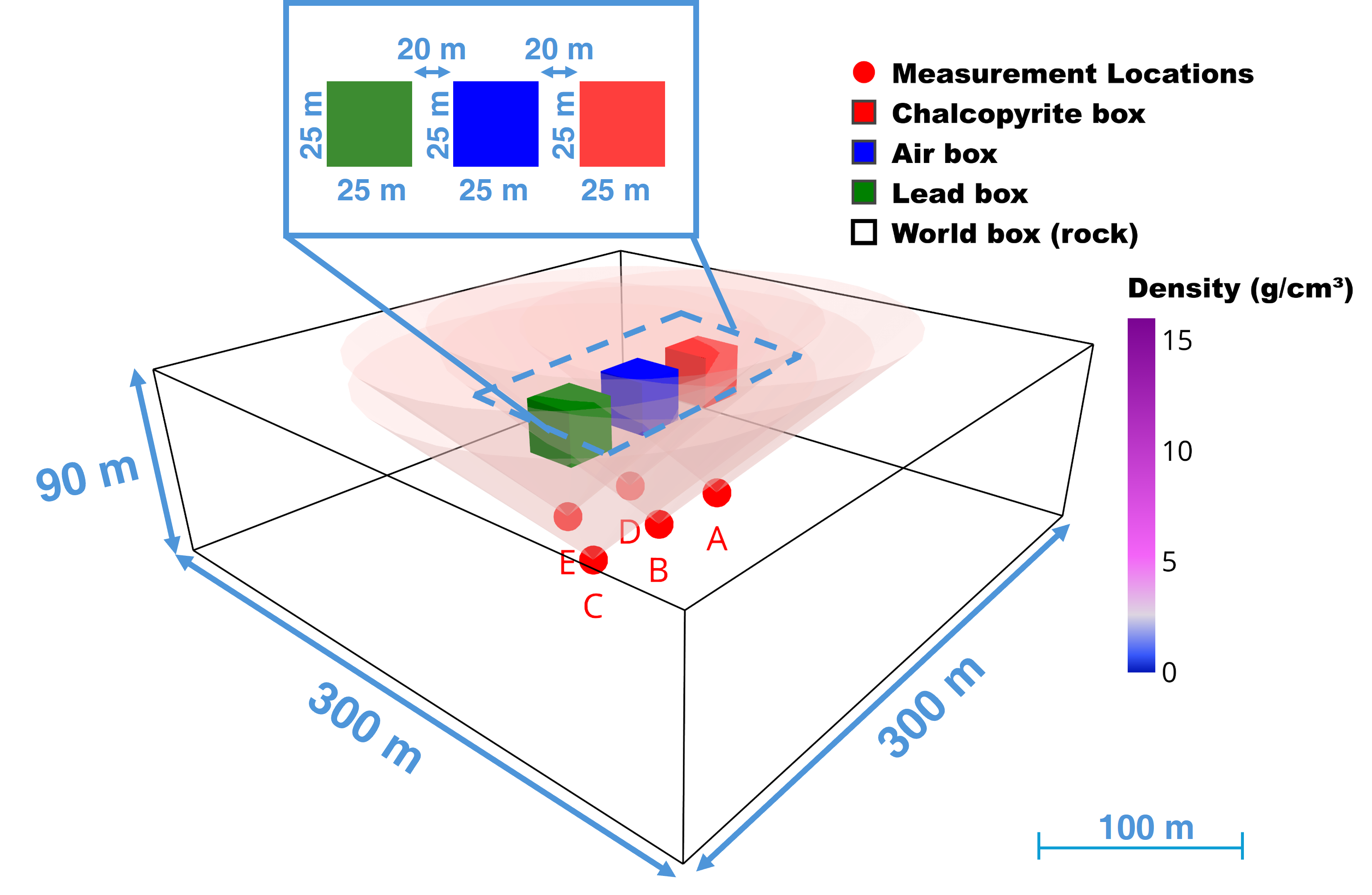}    

    \caption{Overview of the Monte Carlo simulation scenario. Red points indicate the locations of the muon detectors, and pale pink cones originating from the detectors represent the acceptance cones. The region bounded by the black solid lines represents the world box containing rock while the green, blue, and red boxes represent the density anomalies containing lead, air, and chalcopyrite, respectively.}
    \label{c3:fig3}
\end{figure*}

\begin{figure*}[htbp]
    \centering    
    \includegraphics[width=0.8\textwidth]{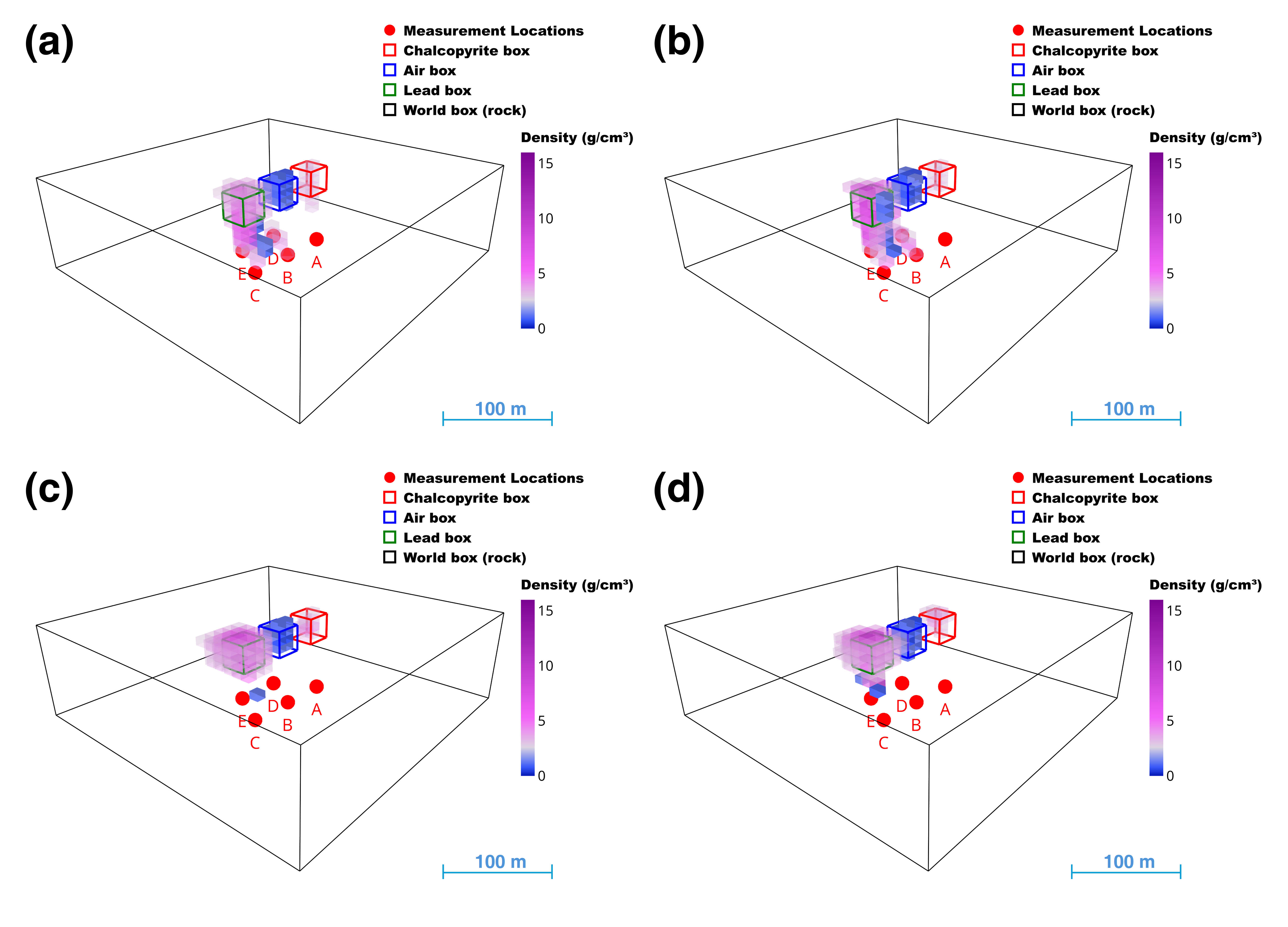}
    \caption{Three-dimensional density reconstruction results with intermediate-density voxels ($1.5$--$3.2\,\mathrm{g/cm^3}$) removed to highlight anomalies. (a) Reconstruction result using the L-BFGS-B algorithm with non-negativity constraint. (b) Reconstruction result using the SART algorithm with non-negativity constraint. (c) and (d) Reconstruction result using the optimized M-H algorithm, initialized with the L-BFGS-B and SART result, respectively.}
    \label{c3:fig4}
\end{figure*}

\begin{figure*}[htbp]
    \centering
    \begin{subfigure}[b]{0.45\textwidth}
        \centering
        \includegraphics[width=\textwidth]{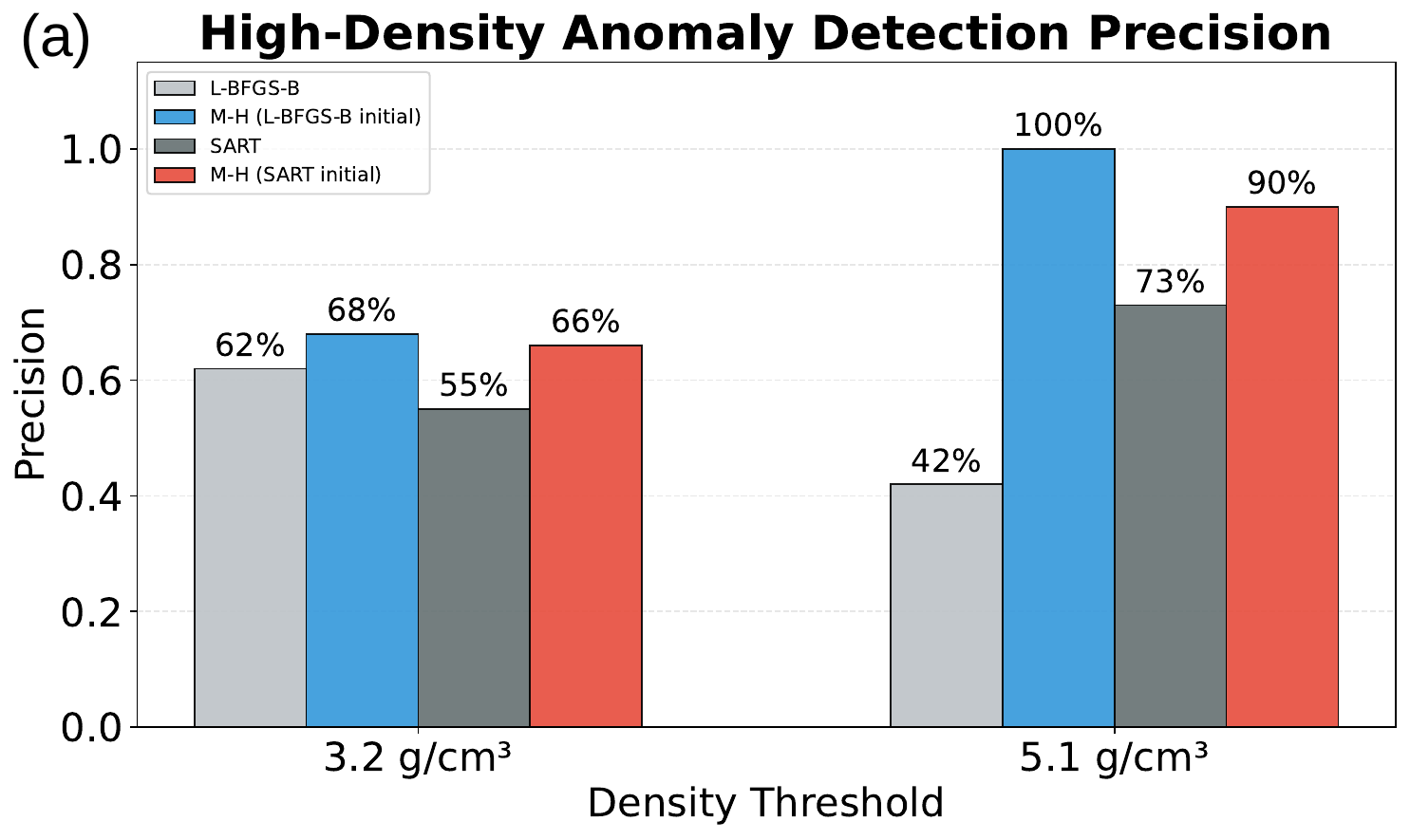}
        \caption{High-density anomaly detection}
        \label{c3:fig10}
    \end{subfigure}
    \hspace{0.04\textwidth}
    \begin{subfigure}[b]{0.45\textwidth}
        \centering
        \includegraphics[width=\textwidth]{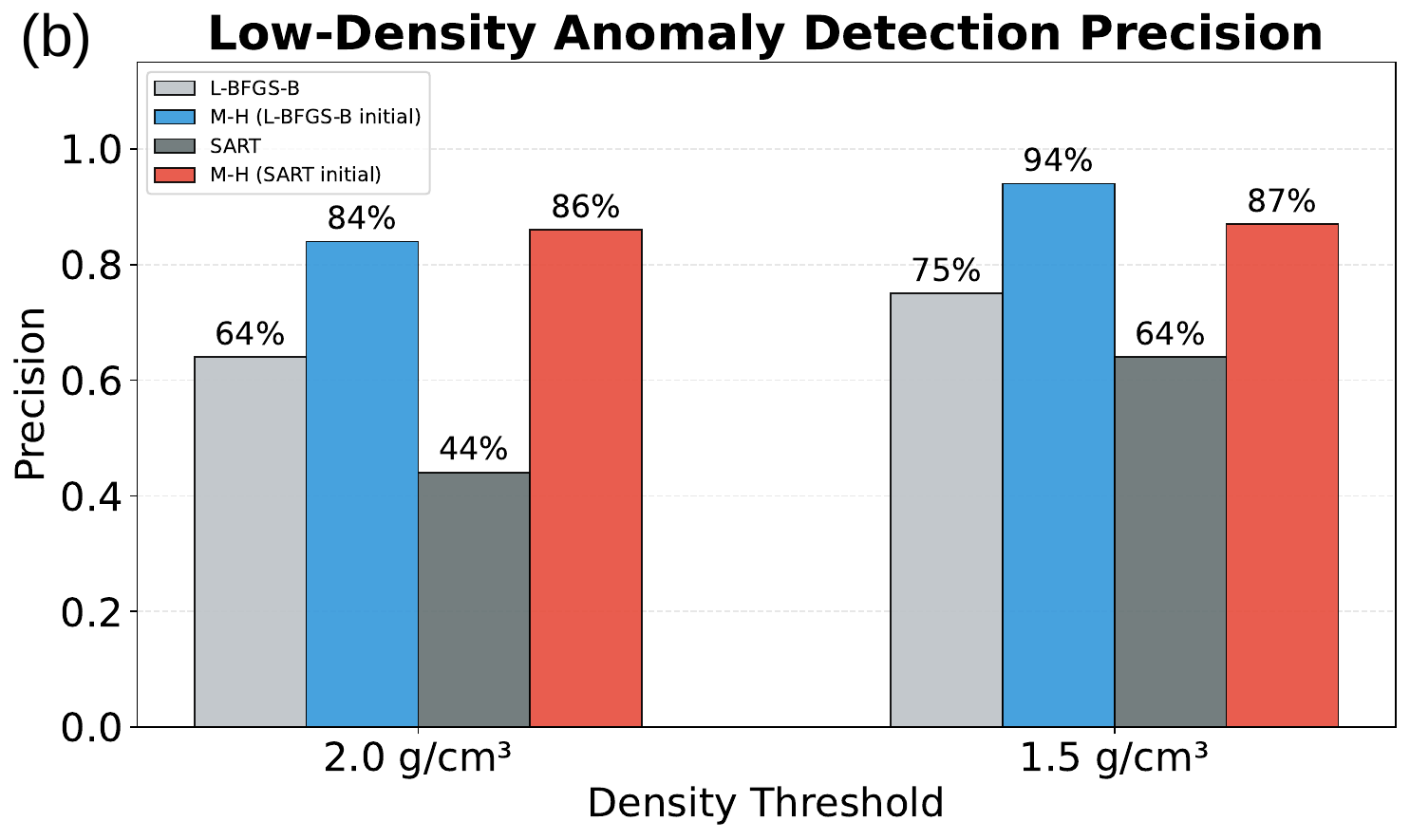}
        \caption{Low-density anomaly detection}
        \label{c3:fig11}
    \end{subfigure}
    
    \caption{Precision of anomaly detection at different density thresholds. Voxels with density above (a) or below (b) the threshold are classified as positive and validated against reference regions: lead (green box) and chalcopyrite (red box) for high-density, air (blue box) for low-density. Blue and red bars represent the optimized M-H algorithm initialized with L-BFGS-B and SART results, respectively, gray bars show baseline performance.}
    \label{c3:fig_precision}
\end{figure*}

\subsection{Capability verification for density reconstruction method by Monte Carlo simulation}
\label{sec:CapVerify}
Before conducting the muography experiment in practice, we first use the Monte Carlo sampling method to test the capability of the optimized M-H algorithm. A scenario is set up to simulate complex underground conditions. As shown in Fig.~\ref{c3:fig3}, five muon detectors are placed inside world box composed of rocks with a uniform density of $2.6 \,\mathrm{g/cm^3}$. The geometric structure of each detector is identical to that of the MuGrid-v2 detector which comprises three layers of plastic scintillators with segmented light guides. Each scintillator layer has a square cross-section of $30\times30\,\mathrm{cm^2}$. The three pale pink cones originating from the detectors represent the acceptance cones with an opening angle of $89^{\circ}$. The world box have dimensions of $300\times300\times90 \,\mathrm{m^3}$. Above these detectors, three boxes containing chalcopyrite (red), air (blue), and lead (green) are placed, with densities of $4.2 \,\mathrm{g/cm^3}$, $1.2\times10^{-3}\,\mathrm{g/cm^3}$ and $11.34 \,\mathrm{g/cm^3}$, respectively. The size of these boxes is $25\times25\times25\,\mathrm{m^3}$. The three boxes are separated by $20\,\mathrm{m}$ from each other and are positioned $50 \,\mathrm{m}$ above the detectors. Here, muon data equivalent to 2 months of observation are generated using a Monte Carlo muon generator EcoMug\cite{paganoEcoMugEfficientCOsmic2021} and used in the subsequent reconstruction.





For comparison, we benchmark our algorithm against L-BFGS-B and SART as baseline algorithms. L-BFGS-B is chosen for its efficiency in handling large-scale constrained optimization problems, while SART represents the standard iterative reconstruction approach in medical tomography. To ensure physical meaningfulness, density values are constrained to be non-negative ($\rho \geq 0\,\mathrm{g/cm^3}$) for all algorithms. Additionally, we use the outputs of these baselines as initialization for our M-H algorithm to demonstrate: (1) the improvement over deterministic methods, and (2) the robustness to different starting points. 

After generating and acquiring muon data of five measurement locations with EcoMug and Geant4 toolkit\cite{agostinelli2003Geant4aSimulation, paganoEcoMugEfficientCOsmic2021}, we reconstruct three-dimensional density distributions using baseline algorithms and optimized M-H algorithm, which are shown in Fig.~\ref{c3:fig4}. The areas outlined by colored lines (green, blue, and red) represent the true locations of the density anomalies. To better visualize the reconstructing density anomaly voxels, we exclude voxels with densities close to the background density of $2.6 \,\mathrm{g/cm^3}$ (specifically, those between $1.5 \,\mathrm{g/cm^3}$ and $3.2 \,\mathrm{g/cm^3}$). 

Figures~\ref{c3:fig4} (a) show the results reconstructed by the L-BFGS-B method with non-negativity constraint. Although all three density anomalies are identified, a large artifact appears beneath the lead box, making it difficult to accurately determine its position. This artifact is primarily caused by an inherent limitation of the reconstruction method. 

Figures~\ref{c3:fig4} (c) show the results reconstructed by the optimized M-H algorithm with bias rate $b=0.2$ and acceptance rate $B_{p}=0.1$, using the previous L-BFGS-B result as the initial sample. Compared with the L-BFGS-B results, the artifact problem is partially mitigated. The artifact beneath the chalcopyrite box completely disappears, and the one beneath the lead box shrinks significantly, making it easier and more accurate to identify the density anomalies.

Figures~\ref{c3:fig4} (b), (d) shows the density distribution reconstructed by the SART algorithm with non-negativity constraint and the optimized M-H algorithm using the SART result as the initial sample, respectively. Similar to the L-BFGS-B case, the M-H algorithm significantly reduces artifacts: the reconstructed anomalies exhibit sharper boundaries and improved spatial localization.

To evaluate reconstruction quality, we select several density thresholds to calculate reconstructed density anomalies precision. For high-density anomaly detection (lead and chalcopyrite), voxels with density above the threshold are classified as positive. For low-density anomaly detection (air cavity), voxels below the threshold are classified as positive. Positive voxels within the known anomaly regions are counted as true positives (TP), while those outside the regions are counted as false positives (FP). Precision is calculated as follow:

\begin{eqnarray}
    P = \frac{\text{TP}}{\text{TP} + \text{FP}}
    \label{c3:eq5}.
\end{eqnarray}

Figures~\ref{c3:fig_precision} (a) and~\ref{c3:fig_precision} (b) quantify the precision of different reconstruction algorithms across anomaly types. The optimized M-H algorithm demonstrates substantial improvements over baseline algorithms: from $42\%$ to $100\%$ in high-density detection at $5.1\,\mathrm{g/cm^3}$ threshold (L-BFGS-B to optimized M-H), with consistent $6\%$--$42\%$ increases across other thresholds. Notably, M-H outperforms both initialization schemes (L-BFGS-B and SART), validating the robustness of our Bayesian sampling approach regardless of the starting point.



\begin{figure*}[htbp]
  \centering
  \includegraphics[width=0.8\textwidth]{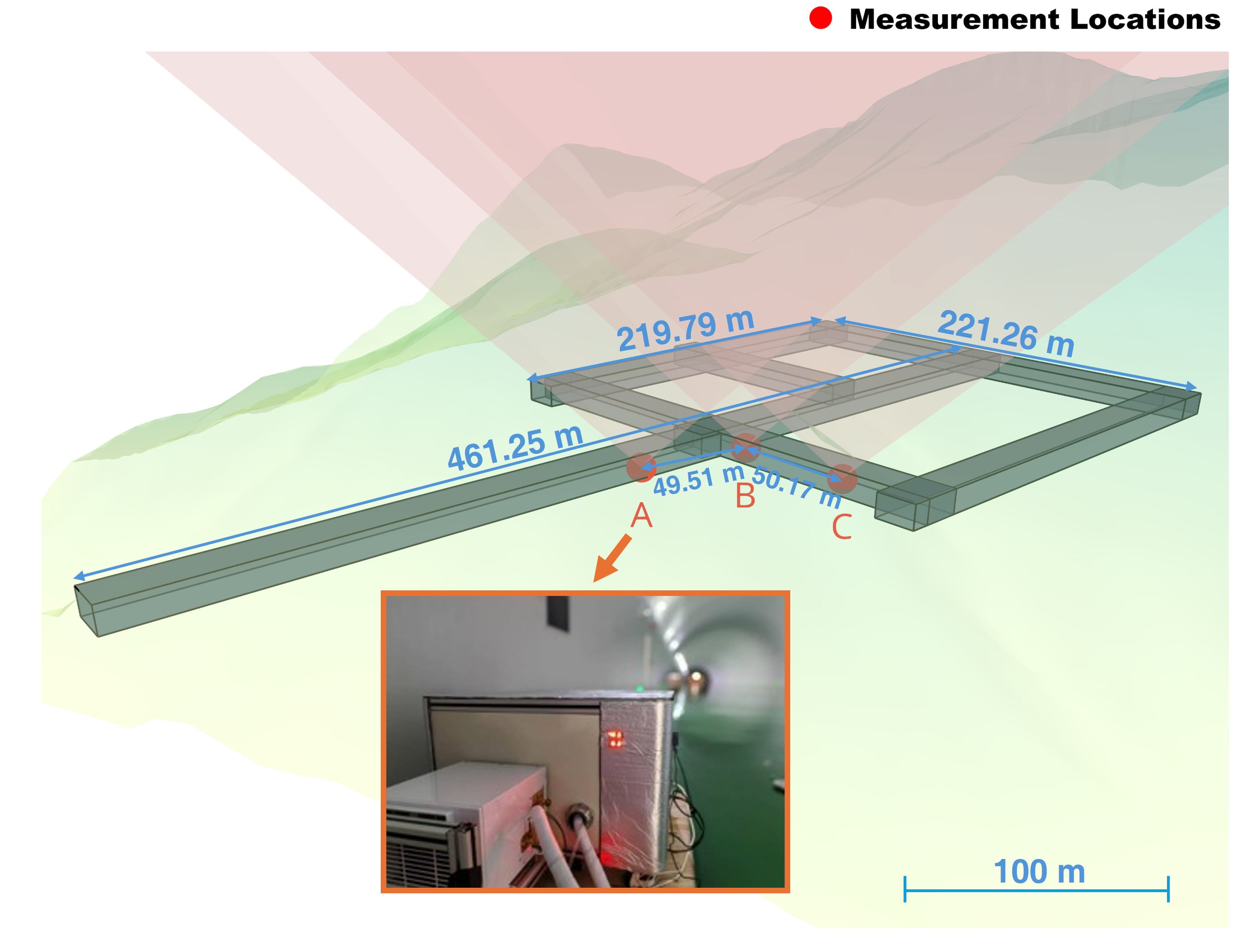}
  \caption{Model of the Tianqin Tunnel with the overlying LiDAR-scanned topography, where the red dots indicate the measurement locations we deployed and the translucent cones show acceptance angles of our detector. Inset: The MuGrid-v2 detector deployed at location A and the central tunnel of the Tianqin Station.}
  \label{c4:fig1}
\end{figure*}

Model of the Tianqin Tunnel with the overlying LiDAR-scanned topography, where the red dots indicate the measurement locations we deployed and the translucent cones show acceptance angles of our detector. Inset: The MuGrid-v2 detector deployed at location A and the central tunnel of the Tianqin Station.

\begin{figure*}[htbp]
    \centering
    \includegraphics[width=0.8\textwidth]{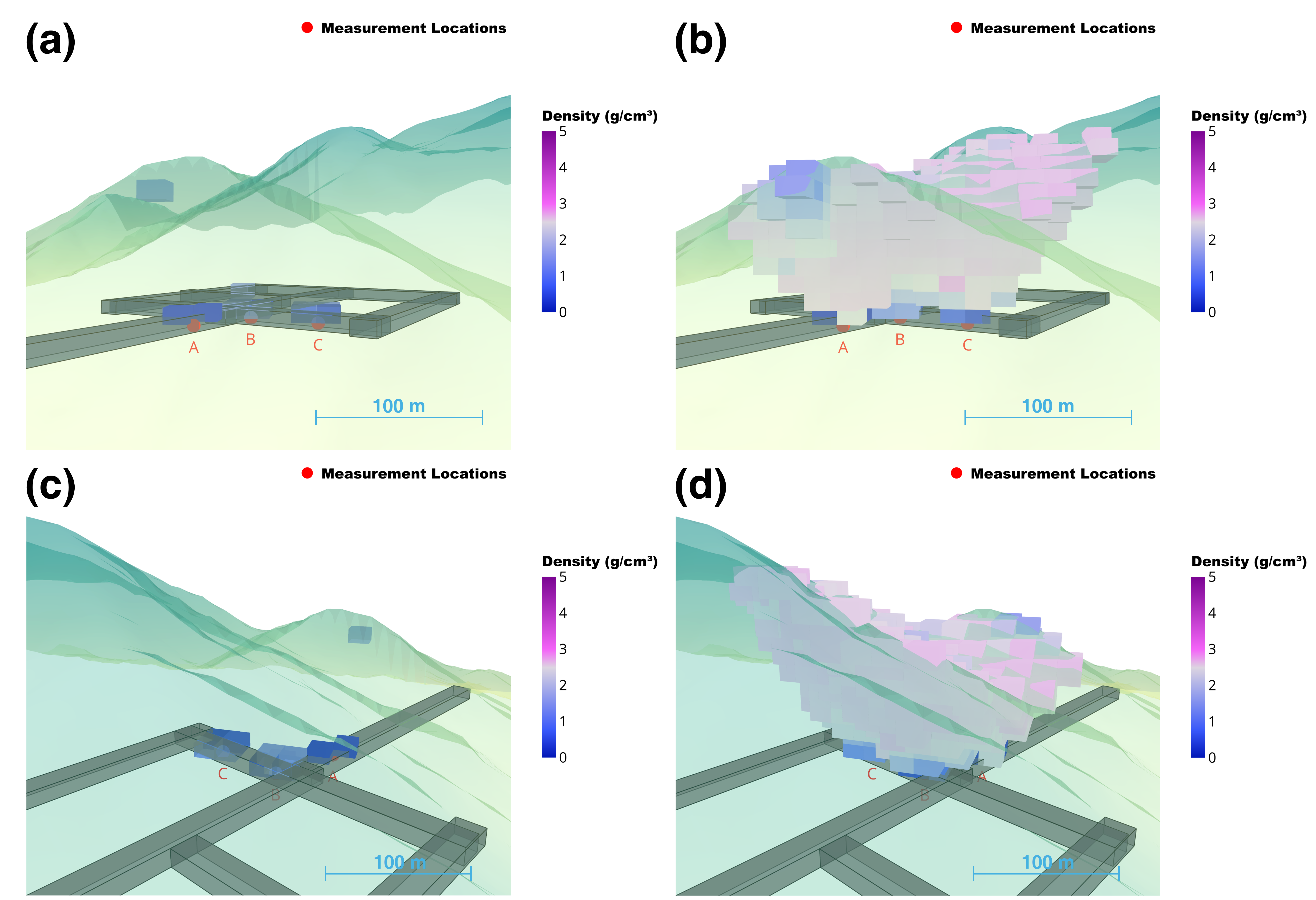}
    \caption{Three-dimensional density reconstruction results at the Tianqin Tunnel. The red dots indicate the measurement locations (points A, B, C). (a),(c) Reconstruction with intermediate-density voxels ($1.5$--$3.2\,\mathrm{g/cm^3}$) removed to highlight anomalies. (b),(d) Full reconstruction showing all voxels.}
    \label{c3:fig5}
\end{figure*}
\begin{figure*}[htbp]
  \centering
  \begin{subfigure}[b]{0.8\textwidth}
    \centering
    \includegraphics[width=\textwidth]{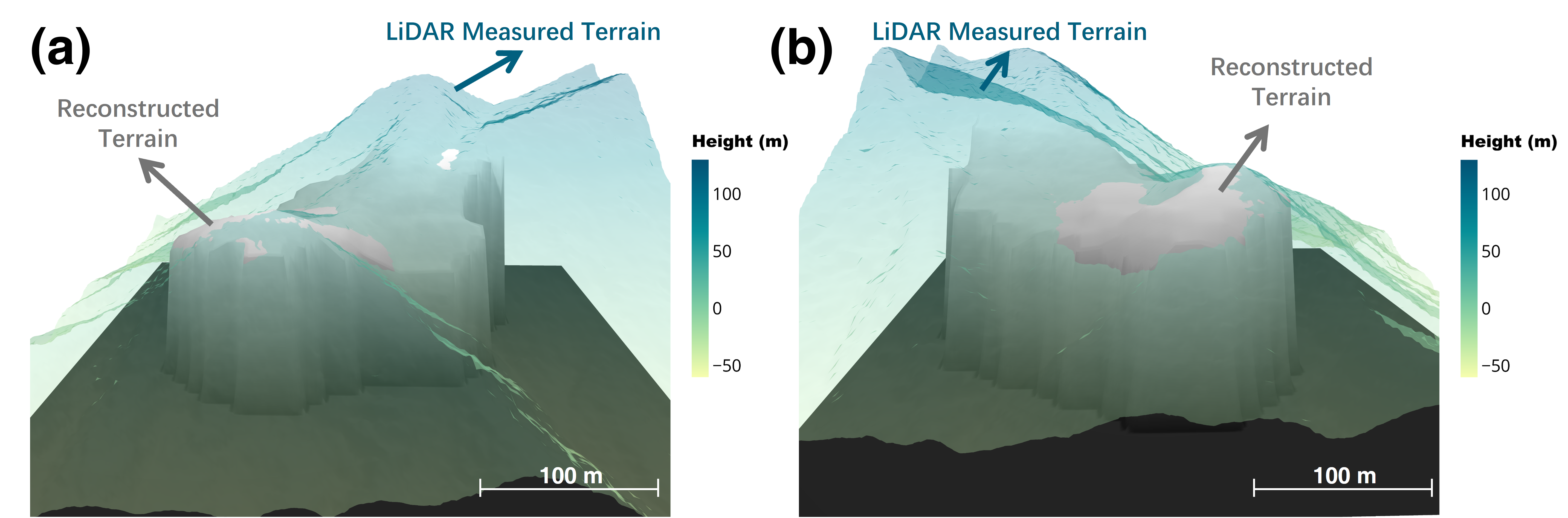}
  \end{subfigure}
  \hfill
  \begin{subfigure}[b]{0.7\textwidth}
    \centering
    \includegraphics[width=\textwidth]{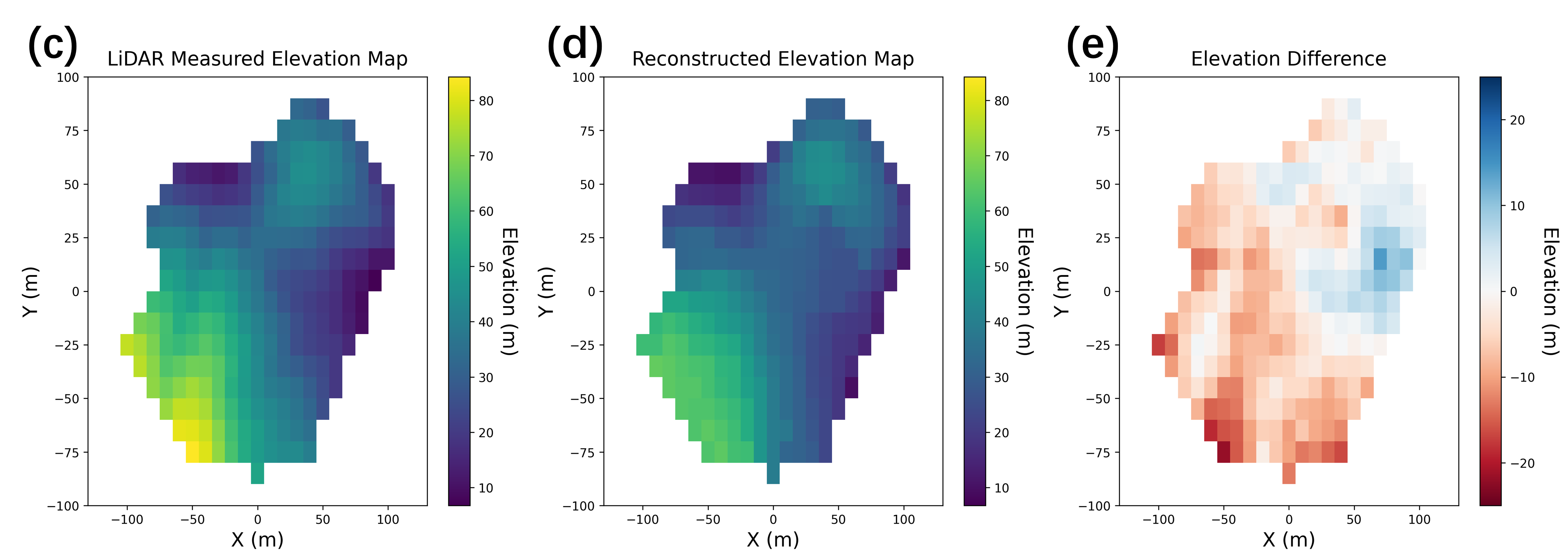}
  \end{subfigure}
  \caption{Air-rock interface reconstruction results at the Tianqin Tunnel. (a)–(b) Different perspectives of the reconstruction air-rock interface, the surface on top is the LiDAR measured elevation map and the profiles below are the reconstructed elevation map. (c) The reconstructed two-dimensional elevation map of the Tianqin Tunnel using experimental data. (d) The two-dimensional elevation map of the Tianqin Tunnel acquired from LiDAR measurements, and (e) Difference between the two elevation maps.}
  \label{c4:fig:interface}
\end{figure*}

\section{Tianqin Tunnel experiment}
\label{DataAc&TTExp}
To further validate the optimized M-H algorithm and benchmark the air-rock interface reconstruction method against LiDAR measurements, we deployed the plastic scintillator muon detector MuGrid-v2 in the Tianqin Tunnel for a field experiment.

\subsection{The MuGrid-v2 detector}
\label{sec:Detector}
Our experiment was conducted with the MuGrid-v2 detector, which is a novel in-house-developed scintillator detector featuring segmented light guides to significantly reduce costs without compromising detection efficiency or spatial resolution\cite{yu2025MuGridv2Novela}. The MuGrid-v2 detector prototype is housed in a steel enclosure measuring approximately $(55\times 80 \times 140)\,\mathrm{g/cm^3}$, with an external air-conditioning unit for thermal management. Inside the steel enclosure there are three detector planes, a data acquisition (DAQ) system, and a temperature and humidity control module. Each detector plane contains two layers of scintillator modules measuring $(30\times 30 \times 1 )\,\mathrm{g/cm^3}$ stacked vertically. 3D-printed optical grid arrays were fabricated and placed on a monolithic plastic scintillator, with 27 wavelength-shifting fibers (WLS) embedded. Since the optical grid segments the scintillator into 27 channels, no addition cutting to the scintillator is required. The distal end of each WLS is coupled to a silicon photomultiplier (SiPM). Signals from the SiPMs are transmitted to the TOFPET2 ASIC evaluation kit accompanied by an self-developed adapter board to achieve analog-to-digital conversion\cite{francesco2016TOFPET2Highperformance}. Finally, motherboard stores the signals in raw ROOT format files for further processing. The MuGrid-v2 detector has an average spatial resolution of $4.6\,mm$ and an angular acceptance of $45^{\circ }$ using a time-based weighted reconstruction algorithm.

\subsection{Experiment setup}
\label{sec:ExpSet}
The Tianqin Geodetic Station, or Tianqin Tunnel, is a high-precision instrumentation platform that belongs to the Space Gravitational Wave Detection Research Program, the Tianqin Project\cite{luo2025ProgressTianQin}. The station is located at Sun Yat-sen University's Zhuhai campus. In addition to its entrance building, most of its tunnels are located beneath the mountain at approximately 100 meters. The tunnel network of the station forms a multi-H-shaped topology, where the long vertical stroke corresponds to the access tunnel. 

We deployed the MuGrid-v2 detector inside the Tianqin Tunnel and collected muons at three different measurement locations as shown in Fig.~\ref{c4:fig1} labeled as points A, B, and C. The semi-transparent colored surface represents the LiDAR-scanned topography above the tunnel. At each measurement location, the detector was oriented in the same direction (with an azimuth angle of $115.3^{\circ}$ and a zenith angle of $0^{\circ}$). The inset picture in Fig.~\ref{c4:fig1} illustrates the typical deployment setup and tunnel environment during the measurement campaign. The measurement locations were arranged to ensure that the overlapping region of all three acceptance cones of the detector could cover the majority of ridges above the tunnel, thereby providing a characteristic topographic features for interface reconstruction. 

Table~\ref{c4:tab1} presents the data acquisition periods for each site. The exposure times were approximately two weeks at sites B and C, one week at site A, and two days for the open sky measurement.




\begin{table}
\caption{Data acquiring period of each site.}
\label{c4:tab1}
\centering
\small  
\begin{tabularx}{\columnwidth}{@{}lXXr@{}}  
\toprule
Detector & Start & End & Effective Data \\
Employed Site & Date & Date & Collecting Time (s) \\
\midrule
A & 01/16/2025 & 01/24/2025 & 648055 \\
B & 01/24/2025 & 02/10/2025 & 1468940 \\  
C & 02/10/2025 & 02/24/2025 & 1468940 \\
Open sky & 02/25/2025 & 02/26/2025 & 57605 \\
\bottomrule
\end{tabularx}
\end{table}

\subsection{Results and discussion}
\label{sec:R&D}
Using the muon data collected at the Tianqin tunnel, we calculate the muon survival rate $S$, from which the minimum energy $E_{\min}$ and opacity $X$ are inferred using the method described in Chapter~\ref{sec:ImageMethod}. Appendixes~\ref{sec:Appen} shows the reconstructed details and data. 

Based on preliminary tests showing that L-BFGS-B initialization yields better performance than SART initialization, we employ L-BFGS-B to solve the inverse problem in Eq.~(\ref{c3:eq1}) and use its solution as the initial sample for the optimized M-H algorithm. The parameters selection was the same as the selection in the previous capability study. The three-dimensional density reconstruction result is shown in Fig.~\ref{c3:fig5}. From (D)-(E) we can infer that except for the blue voxels near the measurement locations (which comprise air inside the tunnel), there are no abnormal voxel clusters exist.

Meanwhile, to reconstruct the air-rock interface at the Tianqin Tunnel, we adopt the uniform rock density $\rho_0 = 2.49 \,\mathrm{g/cm^{3}}$ from geological surveys. To address edge artifacts near the reconstruction boundary and isolated empty pixels due to insufficient data coverage, a median filter is applied to the reconstruction result. Fig.~\ref{c4:fig:interface} (a)-(b) shows the reconstructed result: the top surface represents the air-rock interface obtained from LiDAR measurements, while the vertical profiles below represent the reconstructed air-rock interface.

Fig.~\ref{c4:fig:interface} (c)-(e) shows the two-dimensional elevation maps of air-rock interface reconstructed from muon measurements and LiDAR data together with their difference map. The average error and relative error are $5.121\,\mathrm{m})$ and $16.1\%$, respectively. The errors are primarily attributed to the uncertainty in the uniform density assumption, as in reality the density distribution in the reconstruction region is non-uniform. Despite this simplification, the reconstructed elevation map captures the overall topographic features reasonably well. 




The density reconstruction reveals no significant density anomalies within the detection range. This conclusion is further supported by the good agreement between the reconstructed topography and the LiDAR-measured air-rock interface, as substantial density anomalies would manifest as abnormal depressions or protrusions in the reconstruction. Therefore, the muography measurements at the Tianqin Tunnel indicate no detectable subsurface anomalies above the tunnel structure.

\section{Summary}
\label{sec:Sum}
To conclude, by using traditional three-dimensional reconstruction algorithms (L-BFGS-B and SART) as initial samples for the optimized M-H algorithm, we effectively addressed the artifact generation problem in transmission muography. In validation tests, the density distribution produced by the optimized M-H algorithm exhibits significantly fewer artifacts near anomalies and demonstrates substantial precision improvements over baselines: up to $58\%$ precision improvement in high-density anomaly detection (L-BFGS-B $42\% \to$ optimized M-H $100\%$ at $5.1\,\mathrm{g/cm^3}$ threshold), with consistent $6\%$--$42\%$ increases across other density ranges. Notably, the algorithm shows robust performance regardless of initialization algorithms, validating the effectiveness of our approach. Additionally, we propose an air-rock interface reconstruction method to generate elevation maps when the density distribution is known.

In this paper, we conducted the muography experiment in the Tianqin Tunnel using our in-house-developed muon detector. Leveraging previous geological surveys and LiDAR measurements, we reconstructed both the three-dimensional density distribution and the air-rock interface above the tunnel. The uniform rock density from geological surveys was used to reconstruct the interface, and the result agrees well with the LiDAR-measured elevation map. For the reconstructed density distribution, we found no significant density anomalies above the Tianqin Tunnel. 

This work advances muography in two key aspects. First, we demonstrate that the optimized M-H algorithm significantly reduces artifacts near density anomalies, achieving up to $0.58$ precision improvement over baseline algorithms. Second, when density distribution is known, our air-rock interface reconstruction method provides a complementary tool for geological exploration. The Tianqin Tunnel experiment validates both the capabilities of MuGrid-v2 and the expanded scope of muography applications. However, the current implementation has limitations: the hyperparameters of the optimized M-H algorithm require manual tuning without well-defined selection criteria, and its performance under more complex geological scenarios needs further validation with larger datasets.
\clearpage
\begin{acknowledgments}
This project is supported by Guangdong Basic and Applied Basic Research Foundation under Grant No. 2025A1515010669, the Natural Science Foundation of Guangzhou under Grant No. 2024A04J6243, and Fundamental Research Funds for the Central Universities (23xkjc017) in Sun Yat-sen University. We appreciate the support of School of Geospatial Engineering and Science for the access to the LiDAR-measured elevation map in the Tianqin tunnel.
\end{acknowledgments}


\section*{Data Availability Statement}
The data that support the findings of this study are available within the article


\appendix
\section{Appendixes}
\label{sec:Appen}
\begin{figure*}[b]
    \centering
    \begin{subfigure}[b]{0.2\textwidth}
        \includegraphics[width=\textwidth]{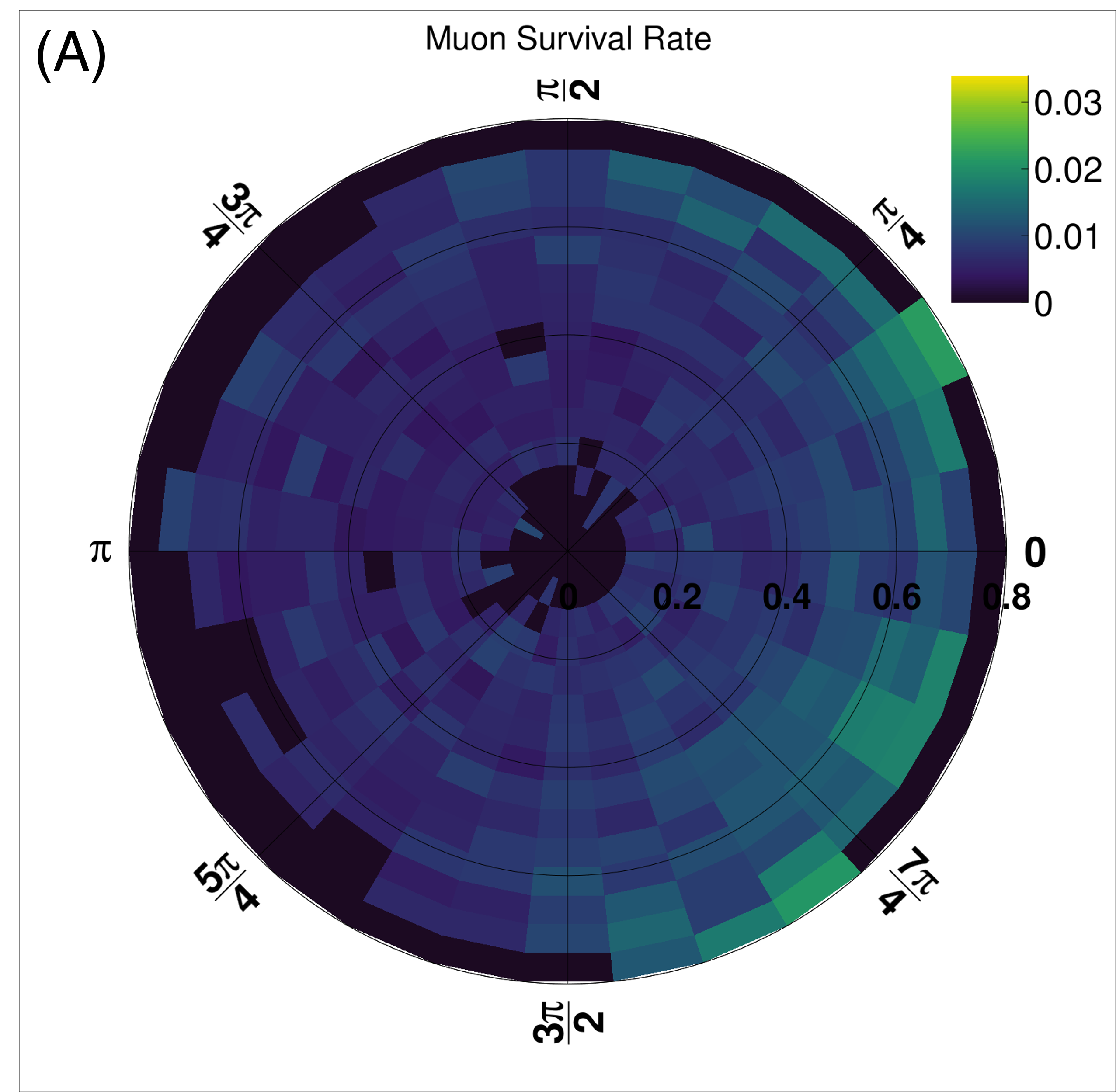}
    \end{subfigure}
    \hspace{0.005\textwidth}
    \begin{subfigure}[b]{0.2\textwidth}
        \includegraphics[width=\textwidth]{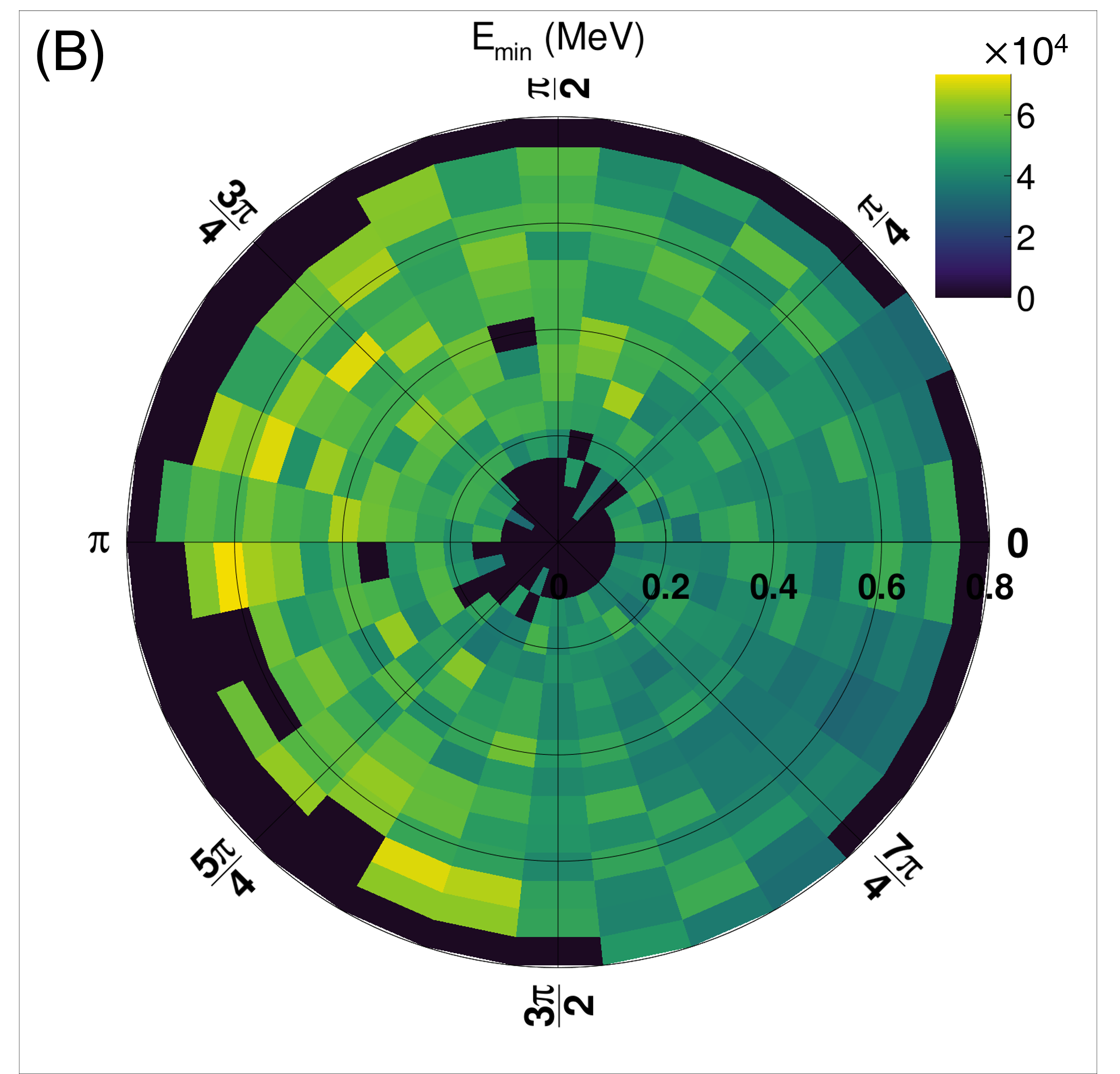}
    \end{subfigure}
    \hspace{0.005\textwidth}
    \begin{subfigure}[b]{0.2\textwidth}
        \includegraphics[width=\textwidth]{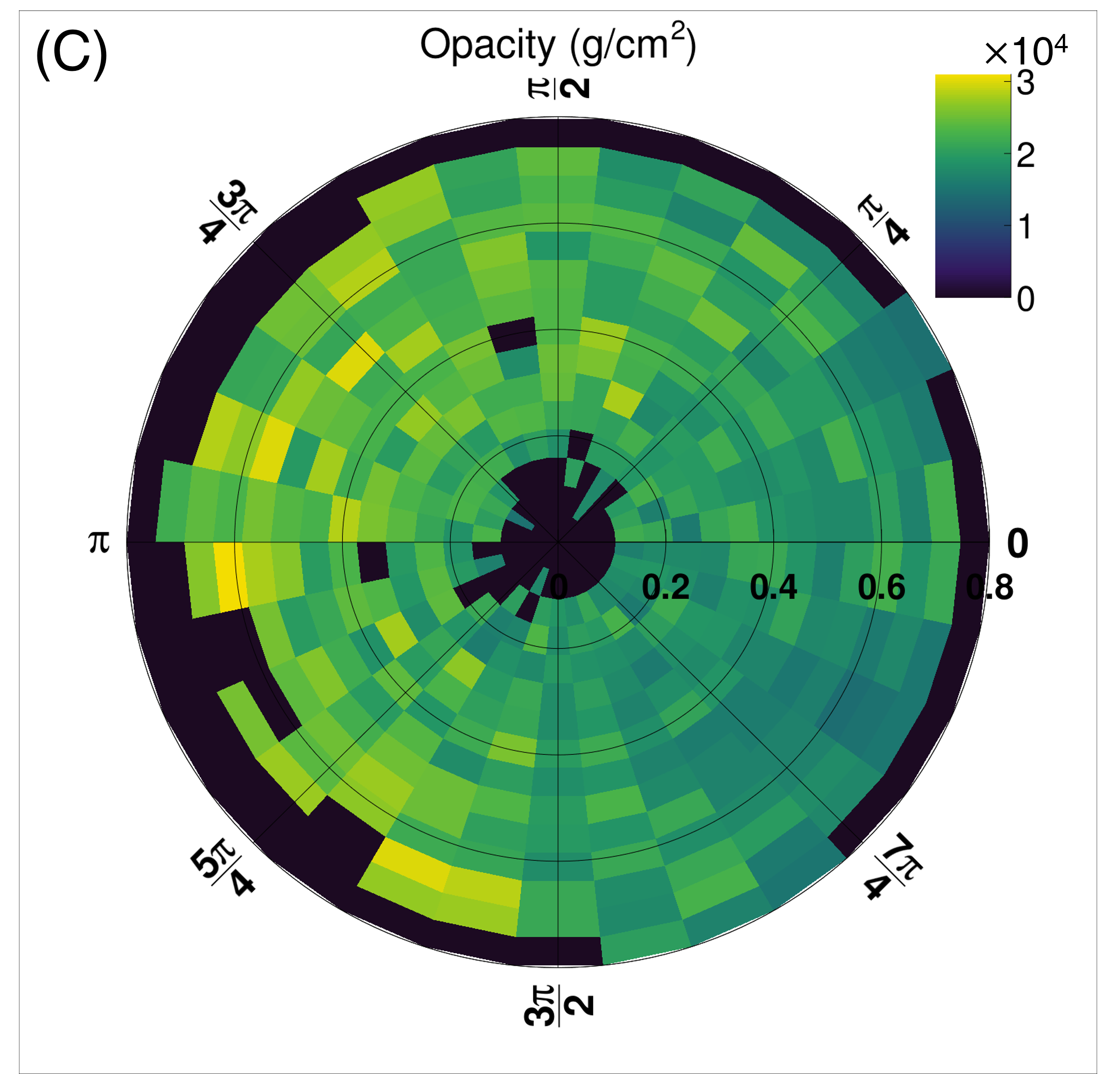}
    \end{subfigure}
  
    \vspace{0.05cm} 
  
    \begin{subfigure}[b]{0.2\textwidth}
        \includegraphics[width=\textwidth]{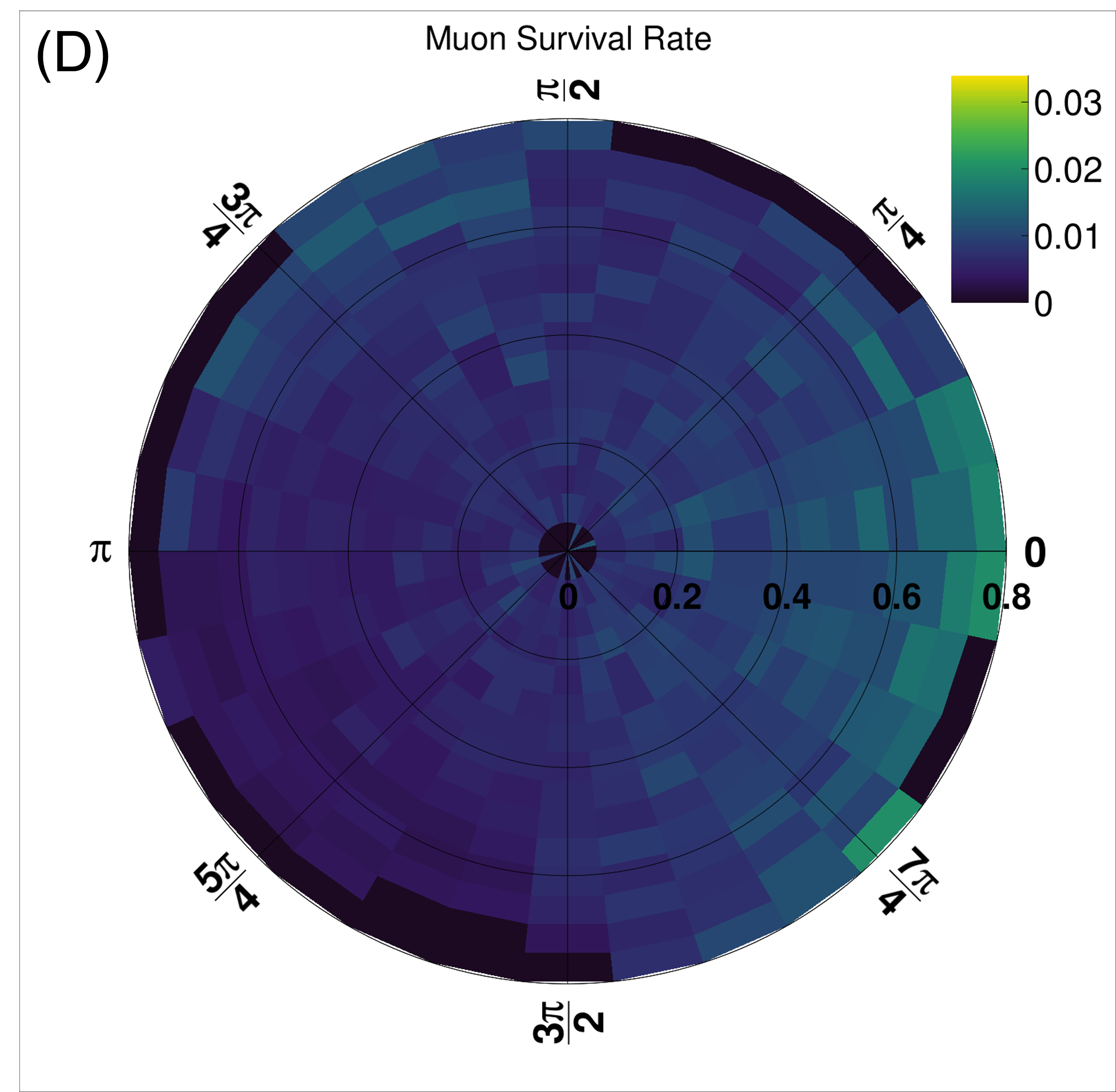}
    \end{subfigure}
    \hspace{0.005\textwidth}
    \begin{subfigure}[b]{0.2\textwidth}
        \includegraphics[width=\textwidth]{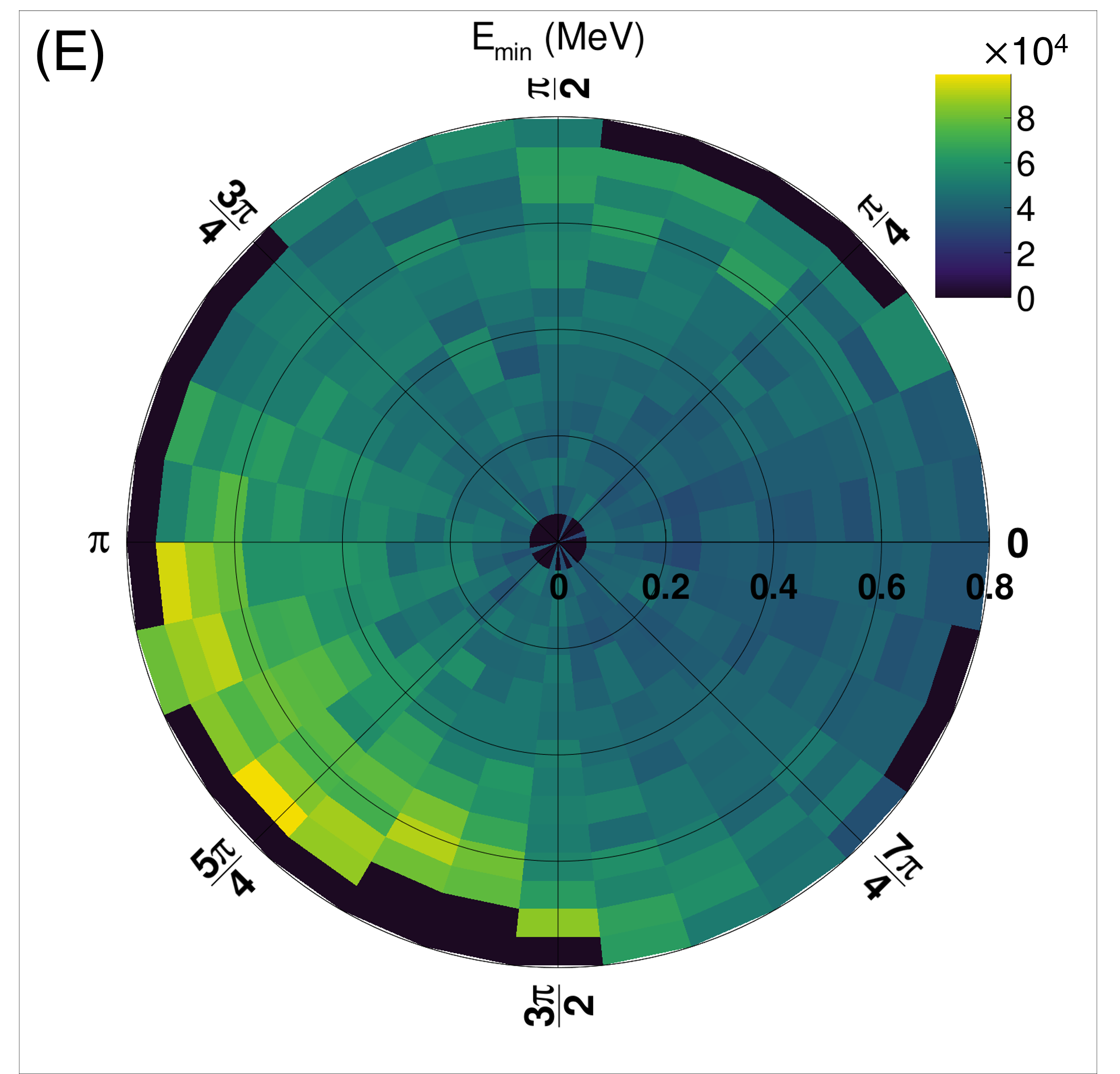}
    \end{subfigure}
    \hspace{0.005\textwidth}
    \begin{subfigure}[b]{0.2\textwidth}
        \includegraphics[width=\textwidth]{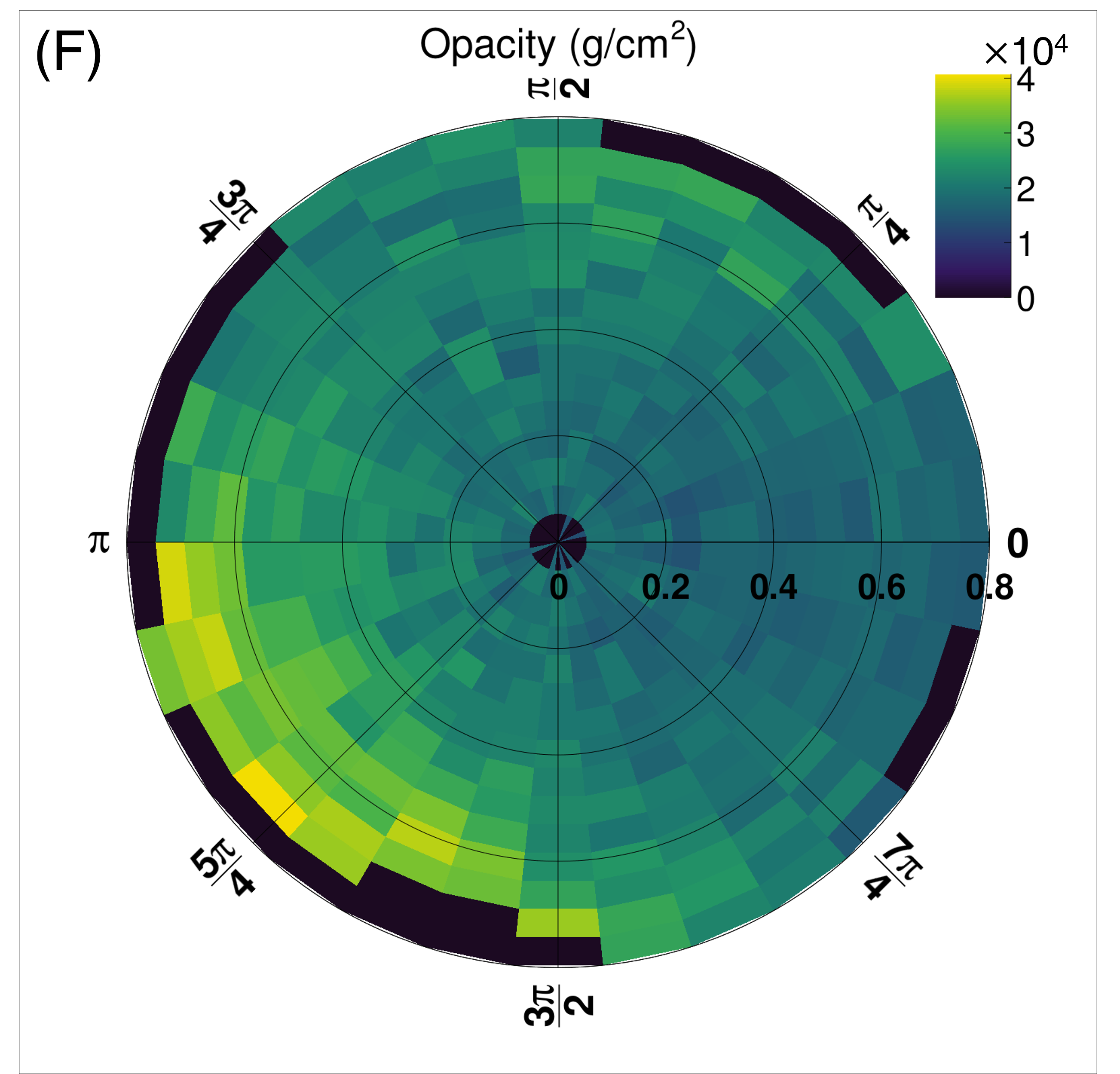}
    \end{subfigure}
  
    \vspace{0.005cm}
  
    \begin{subfigure}[b]{0.2\textwidth}
        \includegraphics[width=\textwidth]{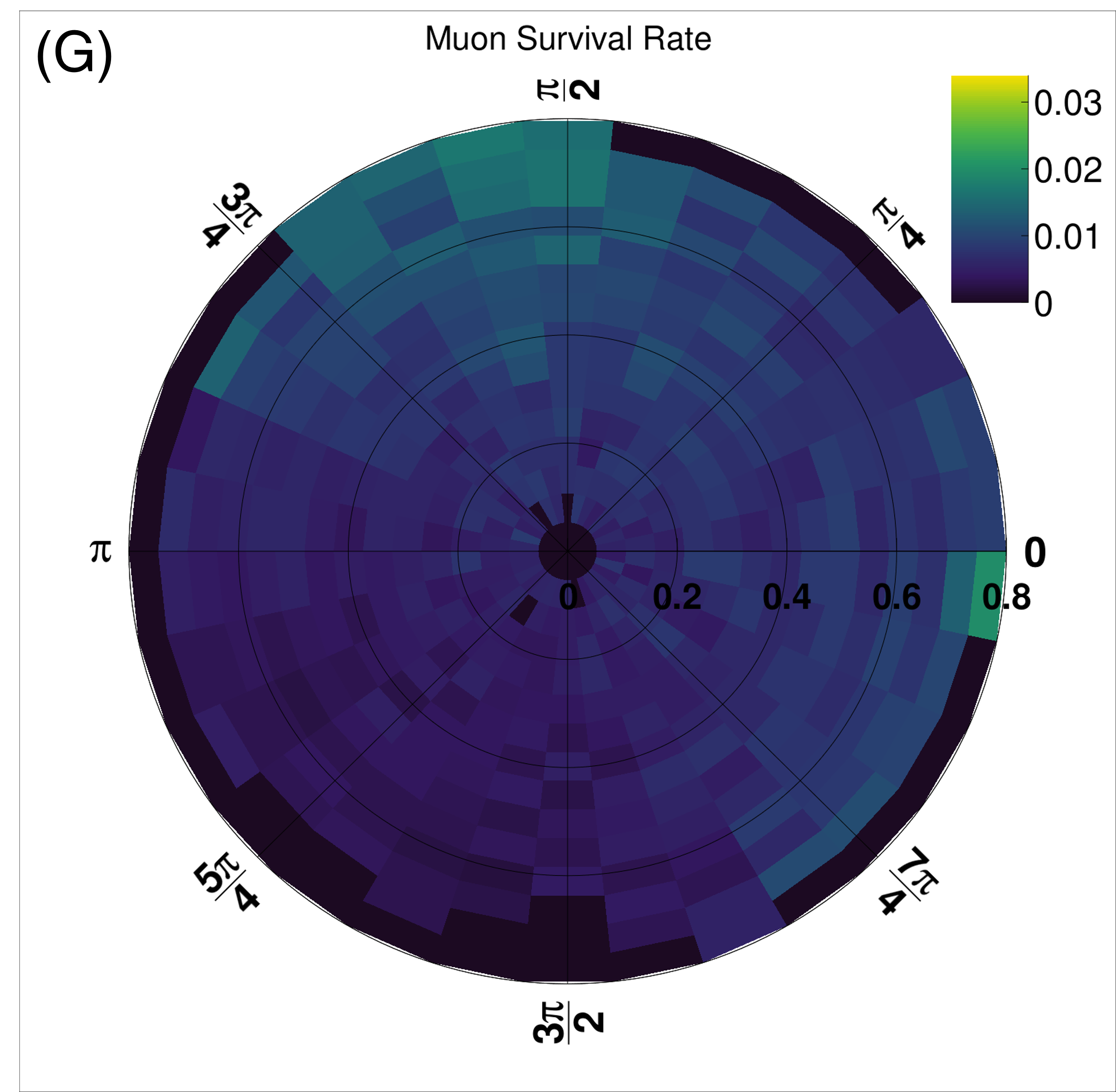}
    \end{subfigure}
    \hspace{0.005\textwidth}
    \begin{subfigure}[b]{0.2\textwidth}
        \includegraphics[width=\textwidth]{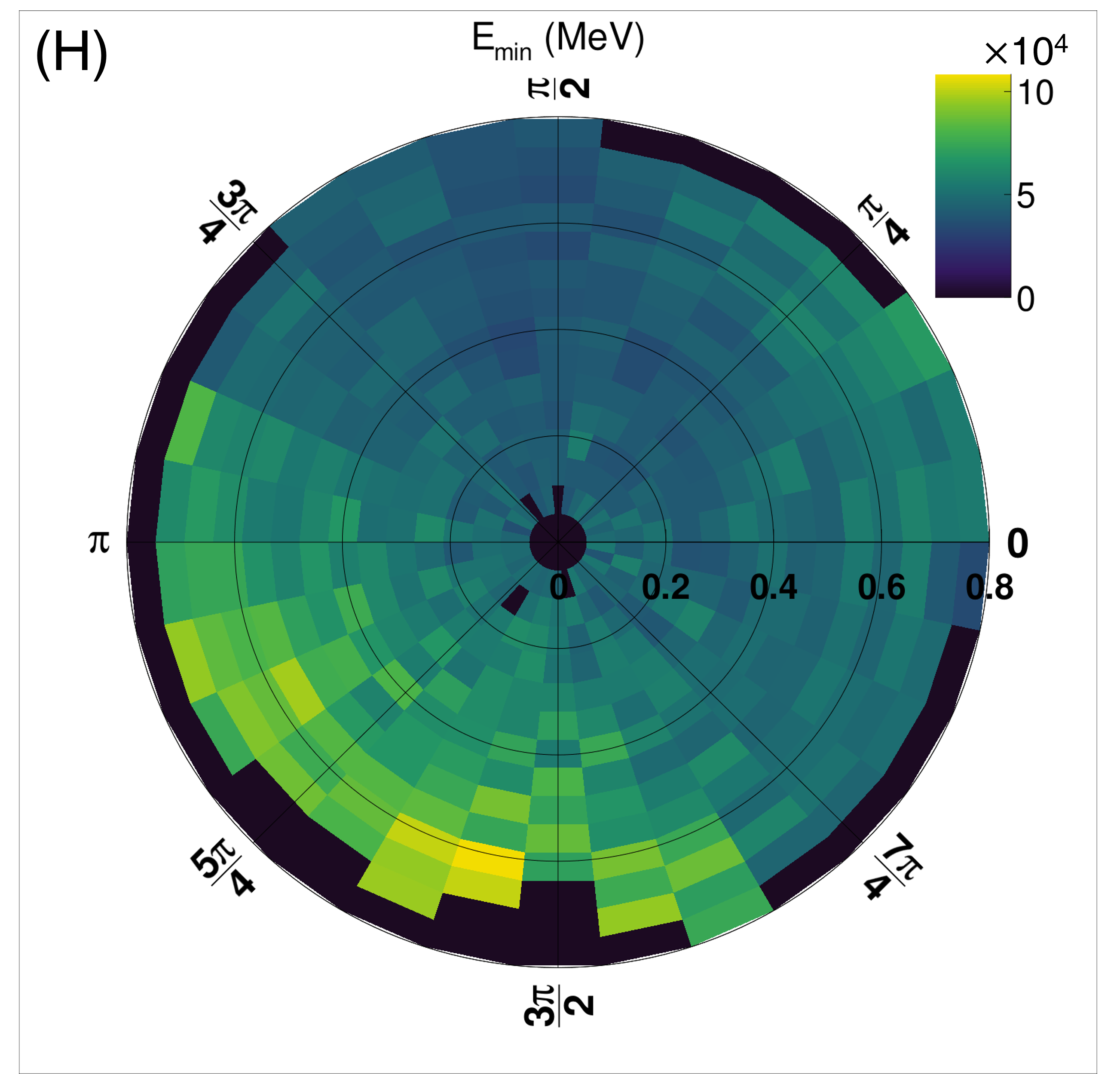}
    \end{subfigure}
    \hspace{0.005\textwidth}
    \begin{subfigure}[b]{0.2\textwidth}
        \includegraphics[width=\textwidth]{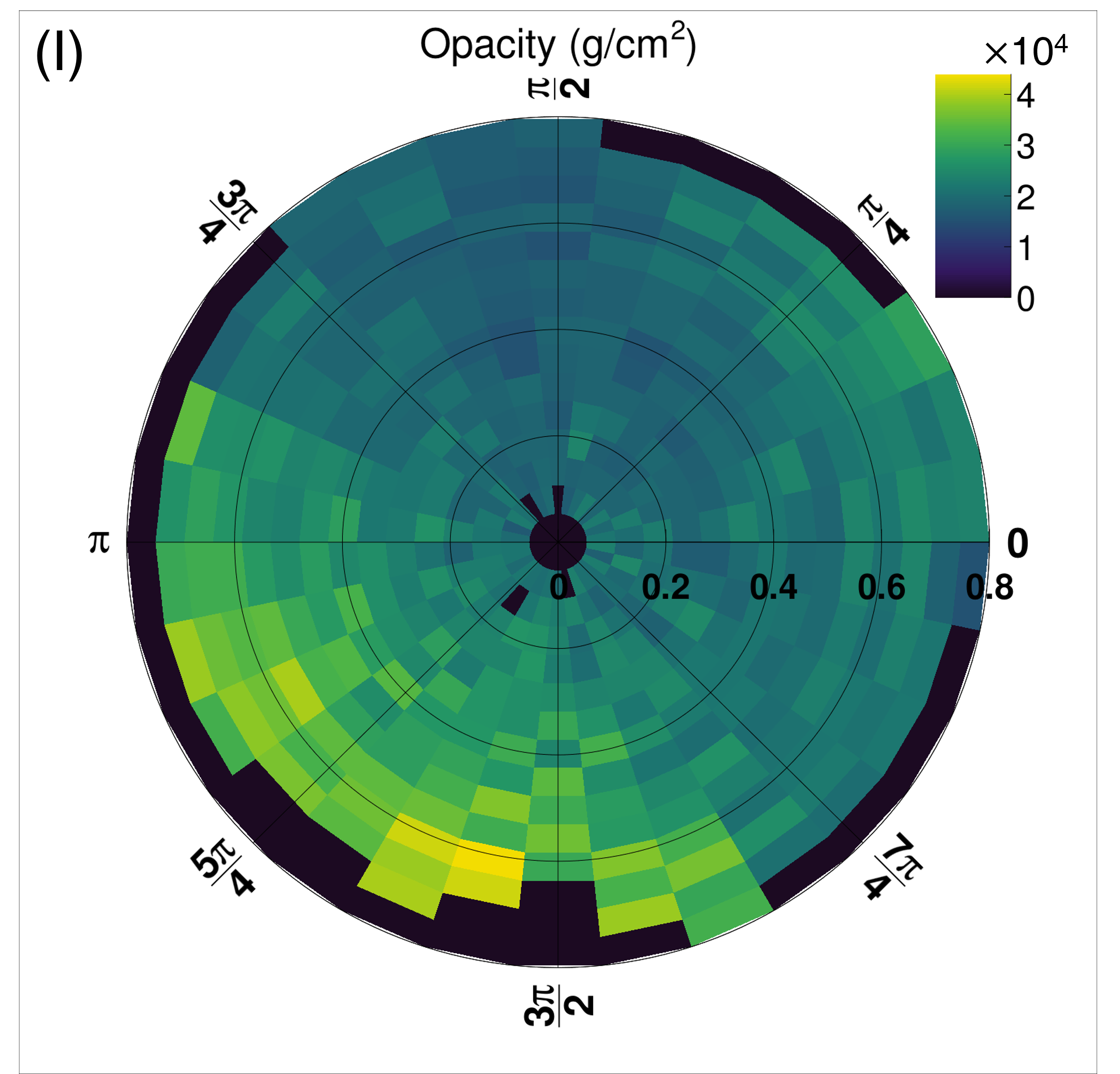}
    \end{subfigure}
  
    \caption{Angular distributions of muon survival rate $S$, minimum energy 
    $E_{\min}$, and opacity $X$ at three measurement locations. 
    (A--C) Location A, (D--F) Location B, (G--I) Location C. 
    The radial axis represents zenith angle (0° to 90°), 
    and the angular axis represents azimuthal angle (0° to 360°).}
    \label{c4:fig3}
\end{figure*}
As described in Chapter~\ref{sec:ImageMethod}, in order to calculate the opacity of different azimuth angles and zenith angles, we need to calculate muon survival rate first. The survival rate can be calculated from received number of muons and data collection times as shown in Eq.~(\ref{c3:eq6}). Then by solving the inverse problem from the equation $E_{min}$ can be acquired. Here we solve it numerically using the FindRoot routine in Wolfram Mathematica\cite{Mathematica14}. Finally we obtain the opacity using the relationship between $E_{min}$ and CSDA range from the PDG database. All these values mentioned above at three measurement locations are shown in Fig.~\ref{c4:fig3}.

As described in Chapter~\ref{sec:ImageMethod}, to calculate the opacity for different azimuth and zenith angles, we first need to determine the muon survival rate. The survival rate can be calculated from the received number of muons and the data collecting time, as shown in Eq.~(\ref{c3:eq6}). Then by solving the inverse problem in Eq.~(\ref{c3:eq6}), we can determine the minimum muon energy $E_{min}$. This inverse problem is solved numerically using the FindRoot routine in Wolfram Mathematica\cite{Mathematica14}. Finally, we obtain the opacity using the relationship between $E_{min}$ and the CSDA range provided by the PDG database. These values for the three measurement locations are shown in Fig.~\ref{c4:fig3}.

\clearpage
\nocite{*}

\begin{thebibliography}{99}
\bibitem{byrdLimitedMemoryAlgorithm1995}Byrd, R., Lu, P., Nocedal, J. \& Zhu, C. A Limited Memory Algorithm for Bound Constrained Optimization. {\em SIAM Journal On Scientific Computing}. \textbf{16}, 1190-1208 (1995)
\bibitem{cimmino3DMuographySearch2019}Cimmino, L., Baccani, G., Noli, P., Amato, L., Ambrosino, F. \& Others 3D Muography for the Search of Hidden Cavities. {\em Scientific Reports}. \textbf{9}, 2974 (2019)
\bibitem{francesco2016TOFPET2Highperformance}Francesco, A., Bugalho, R., Oliveira, L., Pacher, L., Rivetti, A. \& Others TOFPET2: A High-Performance ASIC for Time and Amplitude Measurements of SiPM Signals in Time-of-Flight Applications. {\em Journal Of Instrumentation}. \textbf{11}, C03042 (2016)
\bibitem{guardincerri20173DCosmic}Guardincerri, E. \& Others 3D Cosmic Ray Muon Tomography from an Underground Tunnel. {\em Pure Appl. Geophys.}. \textbf{174}, 2133-2141 (2017)
\bibitem{jourdeExperimentalDetectionUpward2013}Jourde, K., Gibert, D., Marteau, J., De Bremond d'Ars, J., Gardien, S. \& Others Experimental Detection of Upward Going Cosmic Particles and Consequences for Correction of Density Radiography of Volcanoes. {\em Geophysical Research Letters}. \textbf{40}, 6334-6339 (2013)
\bibitem{kak20017Algebraic}Kak, A. \& Slaney, M. 7. Algebraic Reconstruction Algorithms. {\em Principles Of Computerized Tomographic Imaging}. pp. 275-296 (2001)
\bibitem{kak2002PrinciplesComputerized}Kak, A., Slaney, M. \& Wang, G. Principles of Computerized Tomographic Imaging. {\em Medical Physics}. \textbf{29}, 107-107 (2002)
\bibitem{liuDeepInvestigationMuography2024}Liu, G., Yao, K., Niu, F., Li, Z., Tian, H. \& Others Deep Investigation of Muography in Discovering Geological Structures in Mineral Exploration: A Case Study of Zaozigou Gold Mine. {\em Geophysical Journal International}. \textbf{237}, 588-603 (2024)
\bibitem{liuHighprecisionMuographyArchaeogeophysics2023}Liu, G., Luo, X., Tian, H., Yao, K., Niu, F. \& Others High-Precision Muography in Archaeogeophysics: A Case Study on Xi'an Defensive Walls. {\em Journal Of Applied Physics}. \textbf{133}, 014901 (2023)
\bibitem{luo2025ProgressTianQin}Luo, J., Bai, S., Bai, Y., Cai, L., Dang, H. \& Others Progress of the TianQin Project. (arXiv,2025)
\bibitem{mei2021TianQinProject}Mei, J., Bai, Y., Bao, J., Barausse, E., Cai, L. \& Others The TianQin Project: Current Progress on Science and Technology. {\em Progress Of Theoretical And Experimental Physics}. \textbf{2021}, 05A107 (2021)
\bibitem{moralesRemarkAlgorithm7782011}Morales, J. \& Nocedal, J. Remark on “Algorithm 778: L-BFGS-B: Fortran Subroutines for Large-Scale Bound Constrained Optimization”. {\em ACM Trans. Math. Softw.}. \textbf{38}, 7:1-7:4 (2011)
\bibitem{nagamineMethodProbingInnerstructure1995}Nagamine, K., Iwasaki, M., Shimomura, K. \& Ishida, K. Method of Probing Inner-Structure of Geophysical Substance with the Horizontal Cosmic-Ray Muons and Possible Application to Volcanic Eruption Prediction. {\em Nuclear Instruments And Methods In Physics Research Section A: Accelerators, Spectrometers, Detectors And Associated Equipment}. \textbf{356}, 585-595 (1995)
\bibitem{nakamura1986InverseKinematic}Nakamura, Y. \& Hanafusa, H. Inverse Kinematic Solutions With Singularity Robustness for Robot Manipulator Control. {\em Journal Of Dynamic Systems, Measurement, And Control}. \textbf{108}, 163-171 (1986)
\bibitem{nishiyama3DDensityModeling2017}Nishiyama, R., Miyamoto, S., Okubo, S., Oshima, H. \& Maekawa, T. 3D Density Modeling with Gravity and Muon-Radiographic Observations in Showa-Shinzan Lava Dome, Usu, Japan. {\em Pure And Applied Geophysics}. \textbf{174}, 1061-1070 (2017)
\bibitem{nishiyamaBedrockSculptingActive2019}Nishiyama, R., Ariga, A., Ariga, T., Lechmann, A., Mair, D. \& Others Bedrock Sculpting under an Active Alpine Glacier Revealed from Cosmic-Ray Muon Radiography. {\em Scientific Reports}. \textbf{9}, 6970 (2019)
\bibitem{nishiyamaFirstMeasurementIcebedrock2017}Nishiyama, R., Ariga, A., Ariga, T., Käser, S., Lechmann, A. \& Others First Measurement of Ice-Bedrock Interface of Alpine Glaciers by Cosmic Muon Radiography. {\em Geophysical Research Letters}. \textbf{44}, 6244-6251 (2017)
\bibitem{nishiyamaIntegratedProcessingMuon2014}Nishiyama, R., Tanaka, Y., Okubo, S., Oshima, H., Tanaka, H. \& Maekawa, T. Integrated Processing of Muon Radiography and Gravity Anomaly Data toward the Realization of High-resolution 3-D Density Structural Analysis of Volcanoes: Case Study of Showa-Shinzan Lava Dome, Usu, Japan. {\em Journal Of Geophysical Research: Solid Earth}. \textbf{119}, 699-710 (2014)
\bibitem{olive2014ReviewParticle}Olive, K. Review of Particle Physics. {\em Chinese Physics C}. \textbf{38}, 090001 (2014)
\bibitem{paganoEcoMugEfficientCOsmic2021}Pagano, D., Bonomi, G., Donzella, A., Zenoni, A., Zumerle, G. \& Zurlo, N. EcoMug: An Efficient COsmic MUon Generator for Cosmic-Ray Muon Applications. {\em Nuclear Instruments And Methods In Physics Research Section A: Accelerators, Spectrometers, Detectors And Associated Equipment}. \textbf{1014} pp. 165732 (2021)
\bibitem{pCosmicRaysMeasure1955}P, G. Cosmic Rays Measure Overburden of Tunnel. {\em Commonwealth Engineer}. \textbf{455} (1955)
\bibitem{procureur3DImagingNuclear2023}Procureur, S., Attié, D., Gallego, L., Gomez, H., Gonzales, P. \& Others 3D Imaging of a Nuclear Reactor Using Muography Measurements. {\em Science Advances}. \textbf{9}, eabq8431 (2023)
\bibitem{tanakaThreedimensionalComputationalAxial2010}Tanaka, H., Taira, H., Uchida, T., Tanaka, M., Takeo, M. \& Others Three-Dimensional Computational Axial Tomography Scan of a Volcano with Cosmic Ray Muon Radiography. {\em Journal Of Geophysical Research: Solid Earth}. \textbf{115} (2010)
\bibitem{virtanen2020SciPy10}Virtanen, P., Gommers, R., Oliphant, T., Haberland, M., Reddy, T. \& Others SciPy 1.0: Fundamental Algorithms for Scientific Computing in Python. {\em Nature Methods}. \textbf{17}, 261-272 (2020)
\bibitem{wampler1986ManipulatorInverse}Wampler, C. Manipulator Inverse Kinematic Solutions Based on Vector Formulations and Damped Least-Squares Methods. {\em Systems, Man And Cybernetics, IEEE Transactions On}. \textbf{16} pp. 93-101 (1986)
\bibitem{yu2025MuGridv2Novela}Yu, T., Ning, Y., Yuan, Y., Zhao, S., Qi, S. \& Others MuGrid-v2: A Novel Scintillator Detector for Multidisciplinary Applications. {\em Journal Of Applied Physics}. \textbf{138}, 024501 (2025)
\bibitem{zhuAlgorithm778LBFGSB1997}Zhu, C., Byrd, R., Lu, P. \& Nocedal, J. Algorithm 778: L-BFGS-B: Fortran Subroutines for Large-Scale Bound-Constrained Optimization. {\em ACM Trans. Math. Softw.}. \textbf{23}, 550-560 (1997)
\bibitem{borozdin2003RadiographicImaging}Borozdin, K., Hogan, G., Morris, C., Priedhorsky, W., Saunders, A., Schultz, L. \& Teasdale, M. Radiographic Imaging with Cosmic-Ray Muons. {\em Nature}. \textbf{422}, 277-277 (2003)
\bibitem{lechmann2021MuonTomography}Lechmann, A., Mair, D., Ariga, A., Ariga, T., Ereditato, A. \& Others Muon Tomography in Geoscientific Research – A Guide to Best Practice. {\em Earth-Science Reviews}. \textbf{222} pp. 103842 (2021)
\bibitem{lesparre2012DensityMuon}Lesparre, N., Gibert, D., Marteau, J., Komorowski, J., Nicollin, F. \& Coutant, O. Density Muon Radiography of La Soufrière of Guadeloupe Volcano: Comparison with Geological, Electrical Resistivity and Gravity Data. {\em Geophysical Journal International}. \textbf{190}, 1008-1019 (2012)
\bibitem{rosas-carbajal2017ThreedimensionalDensity}Rosas-Carbajal, M., Jourde, K., Marteau, J., Deroussi, S., Komorowski, J. \& Gibert, D. Three-Dimensional Density Structure of La Soufrière de Guadeloupe Lava Dome from Simultaneous Muon Radiographies and Gravity Data. {\em Geophysical Research Letters}. \textbf{44}, 6743-6751 (2017)
\bibitem{borselli2022ThreedimensionalMuon}Borselli, D., Beni, T., Bonechi, L., Bongi, M., Brocchini, D. \& Others Three-Dimensional Muon Imaging of Cavities inside the Temperino Mine (Italy). {\em Scientific Reports}. \textbf{12}, 22329 (2022)
\bibitem{bethe1930theorie}Bethe, H. Zur theorie des durchgangs schneller korpuskularstrahlen durch materie. {\em Annalen Der Physik}. \textbf{397}, 325-400 (1930)
\bibitem{PDGLive2015}Particle Data Group PDG Live: Atomic and Nuclear Properties of Materials.  (2025), http://pdg.lbl.gov/2025/AtomicNuclearProperties/index.html, Accessed: 2026-02-15
\bibitem{Mathematica14}Wolfram Research, I. Mathematica, Version 14.0.  (2024), https://www.wolfram.com/mathematica, Computer software
\bibitem{particledatagroup2018ReviewParticle}Group, P., Tanabashi, M., Hagiwara, K., Hikasa, K., Nakamura, K. \& Others Review of Particle Physics. {\em Physical Review D}. \textbf{98}, 030001 (2018)
\bibitem{agostinelli2003Geant4aSimulation}Agostinelli, S., Allison, J., Amako, K., Apostolakis, J., Araujo, H. \& Others Geant4—a Simulation Toolkit. {\em Nuclear Instruments And Methods In Physics Research Section A: Accelerators, Spectrometers, Detectors And Associated Equipment}. \textbf{506}, 250-303 (2003)
\end{thebibliography}

\end{document}